\documentclass[preprint]{iucr}              

\usepackage{amsmath, amsthm, amssymb, amsfonts, textcomp, gensymb, subfigure, import, color, siunitx, float, bm, graphicx, url}

     \journalcode{S}              

\begin{document}                  



\title{Annealing of focused ion beam damage in gold microcrystals: An \textit{in situ} Bragg coherent X-ray diffraction imaging study}


\author[a]{David}{Yang}
\author[a]{Nicholas}{W. Phillips}
\author[a]{Kay}{Song}
\author[b]{Ross}{J. Harder}
\author[b]{Wonsuk}{Cha}
\cauthor[a]{Felix}{Hofmann}{david.yang@eng.ox.ac.uk  \linebreak Email: felix.hofmann@eng.ox.ac.uk}

\aff[a]{Department of Engineering Science, University of Oxford, Parks Road, Oxford, OX1~3PJ, \country{UK}}
\aff[b]{Advanced Photon Source, Argonne National Laboratory, Argonne, IL 60439, \country{USA}}









\maketitle                        

\begin{abstract}
Focused ion beam (FIB) techniques are commonly used to machine, analyse and image materials at the micro- and nanoscale. However, FIB modifies the integrity of the sample by creating defects that cause lattice distortions. Methods have been developed to reduce FIB-induced strain, however these protocols need to be evaluated for their effectiveness. Here we use non-destructive Bragg coherent X-ray diffraction imaging to study the \textit{in situ} annealing of FIB-milled gold microcrystals. We simultaneously measure two non-collinear reflections for two different crystals during a single annealing cycle, demonstrating the ability to reliably track the location of multiple Bragg peaks during thermal annealing. The thermal lattice expansion of each crystal is used to calculate the local temperature. This is compared to thermocouple readings, which are shown to be substantially affected by thermal resistance. To evaluate the annealing process, we analyse each reflection by considering facet area evolution, cross-correlation maps of displacement field and binarised morphology, and average strain plots. The crystal's strain and morphology evolve with increasing temperature, which is likely to be caused by the diffusion of gallium in gold below $\sim$280\degree C and the self-diffusion of gold above $\sim$280\degree C. The majority of FIB-induced strains are removed by 380-410\degree C, depending on which reflection is being considered. Our observations highlight the importance of measuring multiple reflections to unambiguously interpret material behaviour.
\end{abstract}


\section{Introduction} \label{section:introduction}
Focused ion beam (FIB) techniques utilise a beam of energetic ions, often gallium (Ga) extracted from liquid metal, to remove material at the micro- and nanoscale \cite{Rajput2015}. The ability to control sample features at these previously inaccessible length scales has furthered the understanding of material properties, e.g., by enabling the exploration of size-dependent material behaviour. Most notably, the ``smaller-is-stronger" effect has been extensively studied using nanoindented FIB-machined specimens \cite{Uchic2004,Greer2011}. The deformation behaviour of these samples varied significantly, depending on several competing mechanisms. One mechanism, exhibited by polycrystalline copper films \cite{Kiener2007} and Mo alloy single crystals \cite{Bei2007}, shows that FIB damage can lead to a dense population of defects, making the escape of dislocations at the surface more difficult, which causes an apparent hardening. On the contrary, softening has been reported in FIB-fabricated Mo alloy micropillars with a low initial defect density \cite{Shim2009}. Here FIB-induced dislocation loops act as sources for glide dislocations, leading to a reduction in yield stress \cite{Shim2009}. Additionally, FIB can cause other material changes such as amorphisation \cite{Basnar2003}, local recrystallisation \cite{Rubanov2004}, twin nucleation \cite{Lou2018}, and intermetallic phase formation \cite{Babu2016}, to name a few. 

Experiments on gold (Au) microcrystals have shown that FIB milling can cause lattice strains that extend more than 100 nm below the surface of the material \cite{Hofmann2017a,Hofmann2018a}. This can be a major concern as samples are no longer in their original state and these FIB-induced loads will act in addition to externally applied stresses. Notwithstanding, FIB remains a regularly used tool for 3D material characterisation \cite{Nan2019}, micromechanical test sample fabrication \cite{Brousseau2010}, and top-down preparation of transmission electron microscopy (TEM) samples \cite{Rubanov2004,Rajput2015}. The severity of FIB-induced sample modification depends on specific milling conditions \cite{Hutsch2014}. Methods introduced to mitigate this damage must be versatile and effective to allow FIB milling to reach its full potential as a micromachining tool.

Numerous approaches have been suggested for removing FIB damage: use of a plasma FIB \cite{Ernst2017}, post-milling low energy (2 keV) ion polishing \cite{Giannuzzi2005}, sacrificial coatings \cite{Rubanov2001}, chemical etching \cite{Roediger2011}, and post-manufacture heat treatment \cite{Kiener2012,Lee2016}. Here we investigate the effects of annealing, a process that increases defect mobility and thereby provides a pathway for the removal of FIB-induced defects and, more importantly, the resulting strains. To evaluate the effectiveness of annealing for FIB-induced strain removal, we require a technique that allows nanoscale spatial resolution and high strain resolution. Previous \textit{in situ} TEM studies have been performed, showing that temperatures of half the material's melting point were sufficient to remove FIB-induced dislocation loops in Al micropillars \cite{Lee2016} and reduce defect density in FIB-made Cu samples \cite{Kiener2012}. However, TEM is limited to thin samples, where the close proximity of free surfaces can dominate behaviour. Hence, we require a technique capable of capturing high strain resolution information in 3D micron-sized volumes.

Bragg coherent X-ray diffraction imaging (BCDI) allows for 3D-resolved, nanoscale strain measurements and can provide 3D spatial resolution of less than ten nanometers \cite{Cherukara2018} as well as strain resolution on the order of $\sim$2$\times10^{-4}$ \cite{Carnis2019,Hofmann2020}. BCDI involves fully illuminating a crystalline sample inside the coherent volume of an X-ray beam, which is approximately $1 \mathrm{\mu m}\times1 \mathrm{\mu m}\times1 \mathrm{\mu m}$ at third generation synchrotron sources \cite{Clark2012,Hofmann2017b}. Once the Bragg condition is met for a specific $hkl$ reflection, the diffraction pattern is collected on a pixelated area detector positioned perpendicular to the outgoing wave vector in the Fraunhofer diffraction regime. By rotating the sample about a rocking axis, a 3D coherent X-ray diffraction pattern (CXDP) is collected as the detector sequentially intersects different parts of the chosen 3D Bragg peak in reciprocal space \cite{Robinson2001}. If the CXDP is oversampled by at least twice the Nyquist frequency \cite{Miao2000b}, at least 4 pixels per fringe period, iterative phase retrieval algorithms that apply constraints in real and reciprocal space can be used to recover the phase \cite{Fienup1982,Robinson2001}. The amplitude and phase in reciprocal space are related to the real-space object via an inverse Fourier transform \cite{Miao2000b,Robinson2001,Clark2012} followed by a space transformation from detector conjugated space to orthogonal lab or sample space \cite{Yang2019,Li2019b,Maddali2020}. The resulting amplitude, $\mathbf{\rho(r)}$, where $\mathbf{r}$ is the position vector, is proportional to the effective electron density of the crystalline volume associated with the particular crystal reflection. The phase, $\mathbf{\psi(r)}$, corresponds to the projection of the lattice displacement field, $\mathbf{u(r)}$, onto the Bragg vector, $\mathbf{Q_\mathit{hkl}}$, of the $hkl$ crystal reflection under consideration:

\begin{equation} \label{eq:phase}
    \mathbf{\psi_\mathit{hkl}(r)} = \mathbf{Q_\mathit{hkl}}\cdot\mathbf{u(r)}
\end{equation}

Previously, BCDI has been used to investigate annealing of 3C-SiC nanoparticles \cite{Hruszkewycz2018}, diamond nanoparticles \cite{Hruszkewycz2017b}, magnetite crystals \cite{Yuan2019} and metallic glass \cite{Chen2020}. Furthermore, BCDI has been extensively used to study Au subject to iron diffusion \cite{Estandarte2018}, copper diffusion \cite{Xiong2014a} and catalytic oxidation \cite{Suzana2019}. However, these experiments only considered one reflection, and thus only the strain along the Bragg vector could be resolved. 
Our previous BCDI studies found that FIB-milled Au crystals had large phase features corresponding to lattice dilation and contraction \cite{Hofmann2017b,Hofmann2017a,Hofmann2018a}. These features were concentrated at the top and on the side of the milled crystals--on the surfaces exposed to the FIB beam. Defects such as stair-rod dislocations and their corresponding Burgers vectors were identified in the FIB-affected sample volume \cite{Hofmann2017a,Hofmann2018a}. This contrasts the comparatively flat, homogeneous phase observed in the as-grown crystals \cite{Harder2013,Hofmann2017a,Hofmann2018a}. The study \cite{Hofmann2018a} further characterised the removal of substantial FIB damage by using a 5 keV polishing step. The ability to capture these observations throughout an entire 3D volume with high strain sensitivity and spatial resolution makes BCDI attractive for the study of FIB-damaged materials.

Here we investigate the \textit{in situ} annealing of FIB-milled Au microcrystals using BCDI to examine how the 3D strain and crystal morphology evolve as a function of temperature. We demonstrate the successful measurement of two unique reflections for two separate FIB-milled Au crystals during an \textit{in situ} annealing cycle from 23\degree C (room/ambient temperature) to 587\degree C. After recovering the amplitude and phase of the crystals, we identify temperatures where structural changes occur through observations of crystal facet areas, Pearson's normalised cross-correlation coefficient, $r$, of $\mathbf{u_{\textit{hkl}}(r)}$ and binary morphology masks, and average strain in different crystal regions of interest (ROI), which are defined according to their FIB exposure. The use of two reflections for each sample highlights subtle differences in the lattice relaxation, which is observed to be dependent on the crystallographic direction and the geometry of the sample. We also consider the diffusion of Ga in Au and the self-diffusion of Au to determine which mechanism is probably responsible for specific changes. With four reflections, we faithfully identify temperatures where transitions in crystal shape, $r$, and average strain occur to evaluate the effectiveness of annealing in eliminating FIB-induced material damage.

\section{Experimental methodology} \label{section:experimental_methodology}

\subsection{Sample preparation} \label{subsection:sample_preparation}
Au microcrystals were prepared on a [001]-oriented silicon substrate with a 100 nm thick thermally grown oxide layer. First, the wafer was spin-coated with ZEP resist and patterns of 2 $\mu \mathrm{m}$ wide lines were produced on the substrate using electron-beam lithography. Following the removal of the resist, the wafer was coated with a 3 nm thick Cr adhesion layer followed by 40 nm of Au. A pattern of Au lines remained on the substrate after a ``lift-off" procedure was used to remove the unpatterned areas. Next, the sample was annealed at 1000\degree C in air for 10 h, dewetting the lines to form microcrystals ranging from 200 nm to 1 $\mathrm{\mu m}$ in diameter. Scanning electron microscopy (SEM) was used to select suitable crystals for this study.

\begin{figure} \label{fig:SEM_comparison}
    \centering
    \includegraphics[width=\textwidth,scale=0.5]{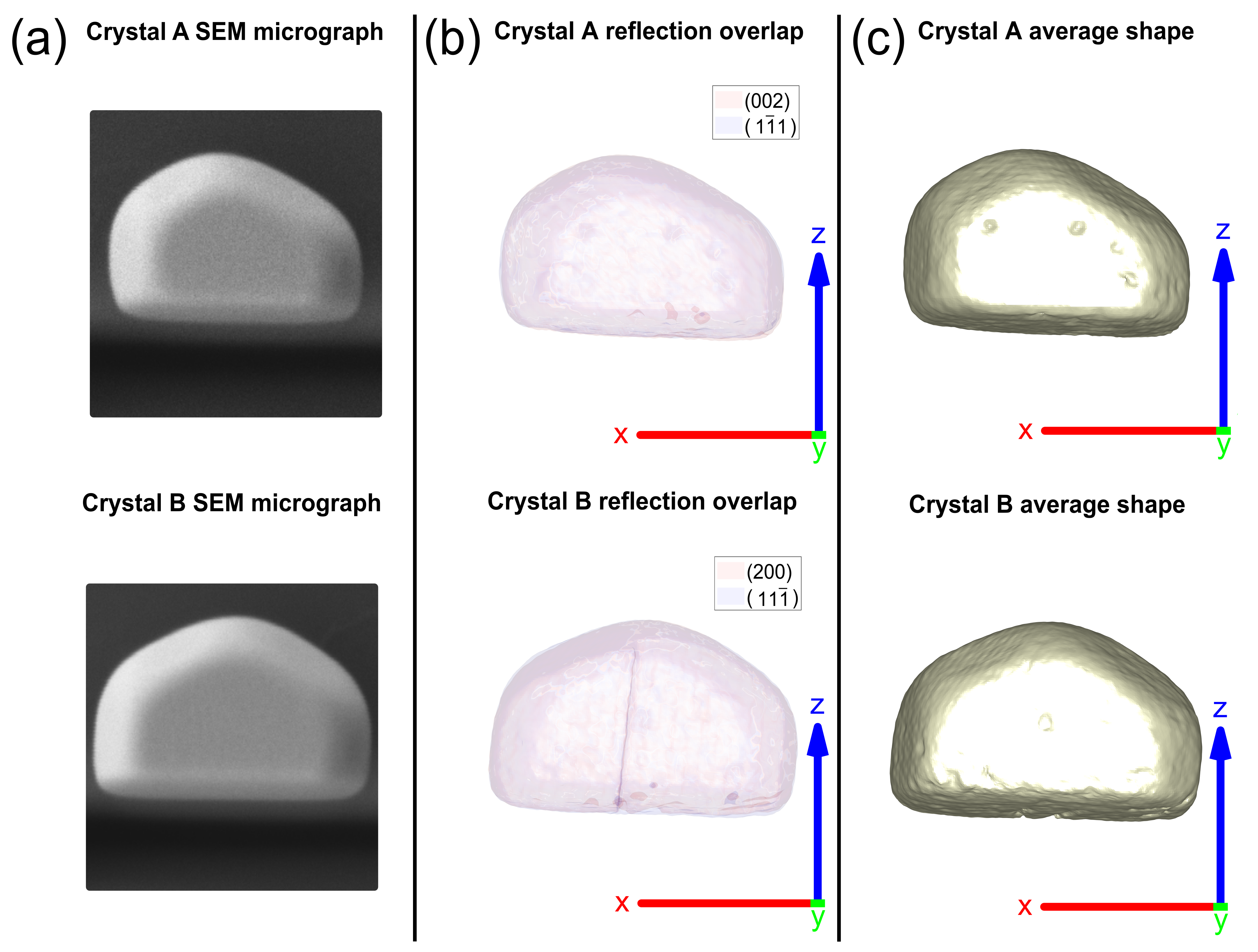}
    \caption{Comparison between SEM micrographs of crystals A and B and sample morphology recovered from BCDI at room temperature (23\degree C). (a) SEM micrographs of the FIB milled crystals. (b) Superimposed reconstructions of the two measured linearly independent reflections for each crystal. Opaque morphologies are rendered to allow examination of their agreement. The near vertical line in crystal B corresponds to a twin domain boundary on the bottom face. (c) Average morphology of the two reflections for each crystal. The amplitude threshold for (b) and (c) is 0.30. The coordinate axes are positioned in sample space and plotted with a length of 500 nm.}
\end{figure}

Using a Zeiss AURIGA FIB-SEM, candidate crystals were subjected to FIB milling with Ga ion energy of 30 keV and beam current of 20 pA, as this closely mimics the final milling cut during the manufacture of micromechanics test specimens \cite{Kiener2008,Li2015,Gong2015}. Approximately one third of the crystal was removed using an incremental trench-milling mode. SEM imaging rather than low dose FIB imaging was used to position the FIB milling scan, as our previous results indicate that even a single FIB image can induce large lattice strains in the Au microcrystals \cite{Hofmann2017a}. Fig. \ref{fig:SEM_comparison}(a) shows high-resolution SEM micrographs of the chosen FIB-milled microcrystals, crystal A and B, in excellent agreement with their respective morphologies recovered from BCDI at room temperature in Fig. \ref{fig:SEM_comparison}(b-c). Crystal A and B were $\sim$10 $\mathrm{\mu m}$ apart on the substrate.

\subsection{Bragg coherent X-ray diffraction} 
\label{subsection:bragg_coherent_X-ray_diffraction}
Synchrotron X-ray diffraction measurements were carried out at the Advanced Photon Source (APS), Argonne National Laboratory, USA. Before BCDI measurements, micro-beam Laue diffraction at beamline 34-ID-E was used to determine the lattice orientation of the crystals on the silicon wafer. A procedure for the alignment of BCDI measurements, based on Laue diffraction orientation measurements, is provided elsewhere \cite{Hofmann2017b}.

BCDI measurements at beamline 34-ID-C used an X-ray beam with 9 keV photon energy $(\lambda = 0.138 \mathrm{\ nm})$ and a spot size of $760 \mathrm{\ nm} \times 750 \mathrm{\ nm}$ (full width at half maximum) using Kirkpatrick–Baez mirrors \cite{Kirkpatrick1948}. The Si substrate was clipped onto a heater cell available at the beamline shown in Appendix \ref{appendix:heater}. A CXDP was collected on a $256 \times 256$ pixel quadrant of a Timepix detector (Amsterdam Scientific Instruments) with a GaAs sensor and pixel size, $p$, of $55 \mathrm{\ \mu m} \times 55 \mathrm{\ \mu m}$ positioned at a distance, $D$, of $1.2$ m away from the sample to ensure that the diffraction pattern was oversampled \cite{Sayre1952}. The lower bound for $D$ was determined by $D=\frac{2dp}{\lambda}$, where $d$ is the sample size. The $(002)$ and $(1\bar{1}1)$ reflections were measured for crystal A and the $(200)$ and $(11\bar{1})$ reflections for crystal B at every temperature step. 

At ambient temperature, the $(1\bar{1}1)$ reflection for crystal A was rotated through an angular range of $0.7\degree$ while the remaining reflections occupied a range of $0.6\degree$. For each rocking curve, a slice through the CXDP was captured every $0.004\degree$ with 0.2 s exposure time and 20 accumulations. A total of 175 increments were collected for the $(1\bar{1}1)$ reflection and 150 increments for the other reflections. Each CXDP measurement was repeated multiple times to optimise the signal-to-noise ratio. For crystal A, 8 scans were collected for the $(002)$ reflection and 14 for the $(1\bar{1}1)$ reflection, while for crystal B, 9 scans were collected for the $(200)$ reflection and 10 for the $(11\bar{1})$ reflection at room temperature.

The angular range, number of accumulations, number of repeated scans and exposure time were adjusted at higher temperatures to sufficiently sample the temperature range as discussed in Section \ref{subsubsection:annealing}. From 90\degree C onward, an angular range of $0.7\degree$ with $0.004\degree$ steps was used, i.e. 175 increments were collected in the rocking scan for CXDPs above room temperature, and 10 accumulations were taken instead of 20. Only 3 repeated scans were collected between 90\degree C and 209\degree C inclusive. From 213\degree C onward, a 0.1 s exposure time was used and only 2 scans were measured, with the exception of the $(1\bar{1}1)$ reflection at 213\degree C where 3 scans were measured. After performing flat-field and dead time corrections, repeated scans were aligned using a 3D version of the approach proposed by \cite{Guizar-Sicairos2008} to maximise their cross-correlation and only scans with $r$ greater than 0.975 were averaged to return the CXDP used for phasing. All scans were summed and used for phase retrieval.

\subsubsection{\textit{In situ} annealing} \label{subsubsection:annealing}
\begin{figure} \label{fig:temperature_profile}
    \centering
    \includegraphics[width=\textwidth]{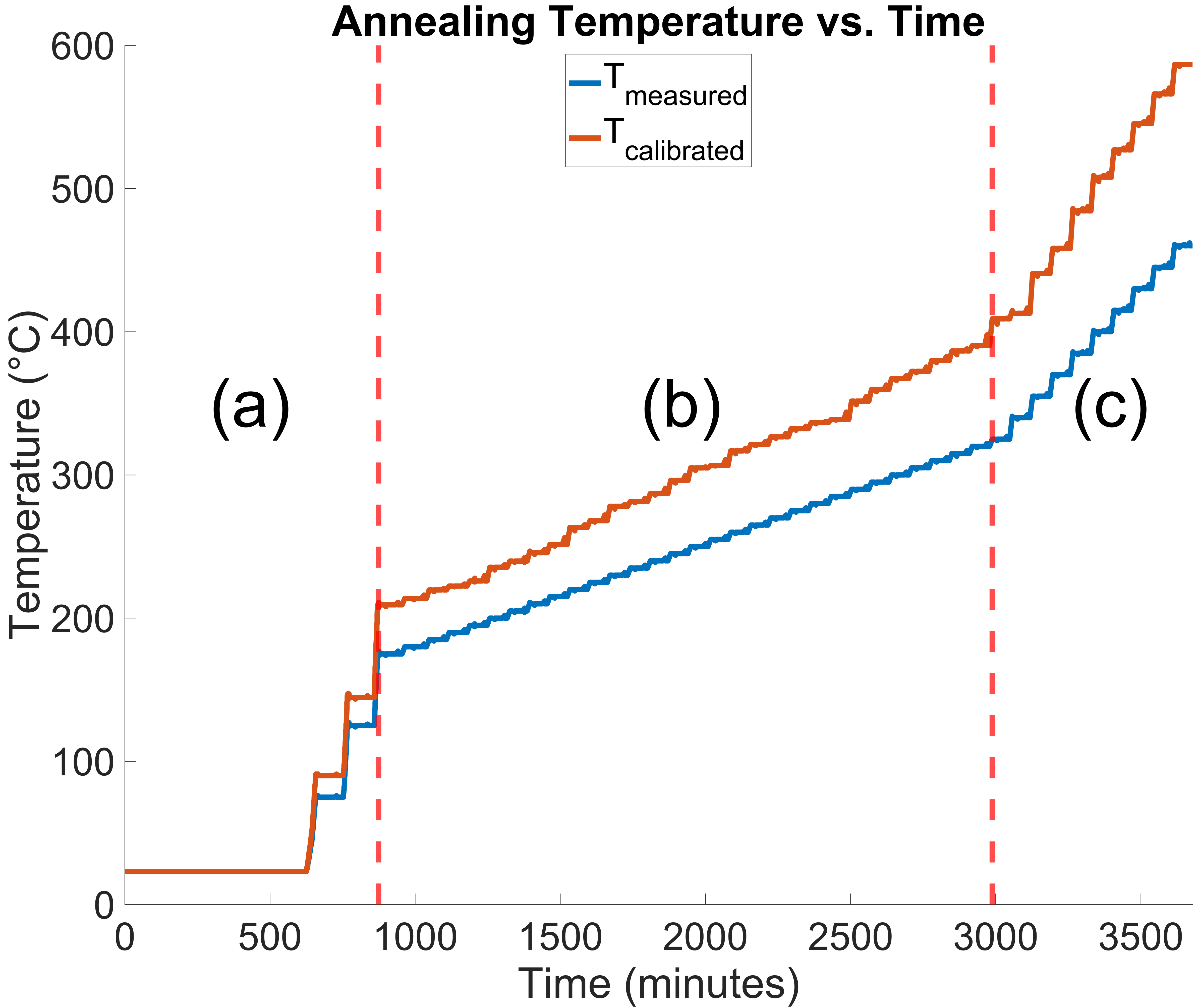} 
    \caption{The annealing temperature profile as a function of time. Changes in the temperature step size are marked by the red dashed lines. (a) $\mathrm{T_{calibrated}}$ was raised from room temperature to 209\degree C using $\sim$50\degree C steps with a ramp rate of 2\degree C/min. (b) From 209\degree C to 409\degree C, $\sim$5\degree C steps were used with a ramp rate of 50\degree C/min. Finally, from 409\degree C to 587\degree C, $\sim$20\degree C steps were used with a ramp rate of 50\degree C/min. The ramp rate reported here is with respect to $\mathrm{T_{measured}}$ and was lower in (a) to ensure that the alignment macro could update the positions of the reflections for larger temperature steps.}
\end{figure}

The temperature was increased using the step and hold temperature profile shown in Fig. \ref{fig:temperature_profile}. Here we note that the calibrated temperature of the Au microcrystals, $\mathrm{T_{calibrated}}$, is different from the temperature recorded using the thermocouple, $\mathrm{T_{measured}}$. This discrepancy arises from the thermal resistance between the thermocouple (attached to the sample clip), and the microcrystals in the heating setup, shown in Appendix \ref{appendix:heater}. To calculate $\mathrm{T_{calibrated}}$ from $\mathrm{T_{measured}}$, we use the samples as local measurement devices. We consider the thermal expansion of pure Au, $\epsilon_\mathrm{thermal}=\alpha_\mathrm{Au}\Delta T$, where $\Delta T$ is the temperature change and $\alpha_\mathrm{Au}$ is the linear thermal expansion coefficient \cite{Howatson2009}, to define $\mathrm{T_{calibrated}}$. We assume that implanted Ga ions after FIB milling have negligible effect on the Au thermal expansion coefficient, as the ions only penetrate 20 nm below the surface \cite{Hofmann2018a}.

Since Au has fcc crystal structure, lattice thermal expansion acts uniformly in all directions, causing the Bragg angle to decrease with increasing temperature. To track the Bragg peak, an alignment macro was run during the temperature ramp that ensured the diffractometer was positioned such that the maximum intensity was always centred in the top left quadrant of the detector immediately before the start of the rocking scans. This protocol ensured that the strain measured by BCDI was relative to the zero-strain reference region for each reflection as explained in Section \ref{subsection:phasing}, allowing a comparison between temperatures. By tracking the change in detector position at every temperature step, $i$, the lattice parameter, $a_i$, at each step can be calculated:

\begin{equation} \label{eq:lattice_constant}
    a_{i} = \frac{\lambda}{2\sin(\theta_i)}\sqrt{h^2+k^2+l^2} \mathrm{,\hspace{2mm} where\hspace{2mm}} \theta_i = \frac{1}{2}\cos^{-1}\left(\frac{\mathbf{S}_{i} \cdot \mathbf{S}_\mathrm{0}}{\mathbf{S}_\mathrm{0} \cdot \mathbf{S}_\mathrm{0}}\right)\\
\end{equation}

where $\theta_i$ is the Bragg angle, $\mathbf{S}_{i}$ is the outgoing wavevector (the detector position) and $\mathbf{S}_\mathrm{0}$ is the incident wavevector. Using $a_{i}$ and the literature value of $\alpha_\mathrm{Au}$, we calculate $\mathrm{T_{calibrated}}$:

\begin{equation} \label{eq:calibrated_temperature}
    T_{\mathrm{calibrated},i} = \frac{\epsilon_{\mathrm{thermal},i}}{\alpha_\mathrm{Au}}+23\degree \mathrm{C,\hspace{2mm} where\hspace{2mm}}\epsilon_{\mathrm{thermal},i} = \frac{a_{i}-a_{23\degree \mathrm{C}}}{a_{23\degree \mathrm{C}}}
\end{equation}

\begin{figure} \label{fig:thermal_strain}
    \centering
    \includegraphics[width=\textwidth,scale=0.5]{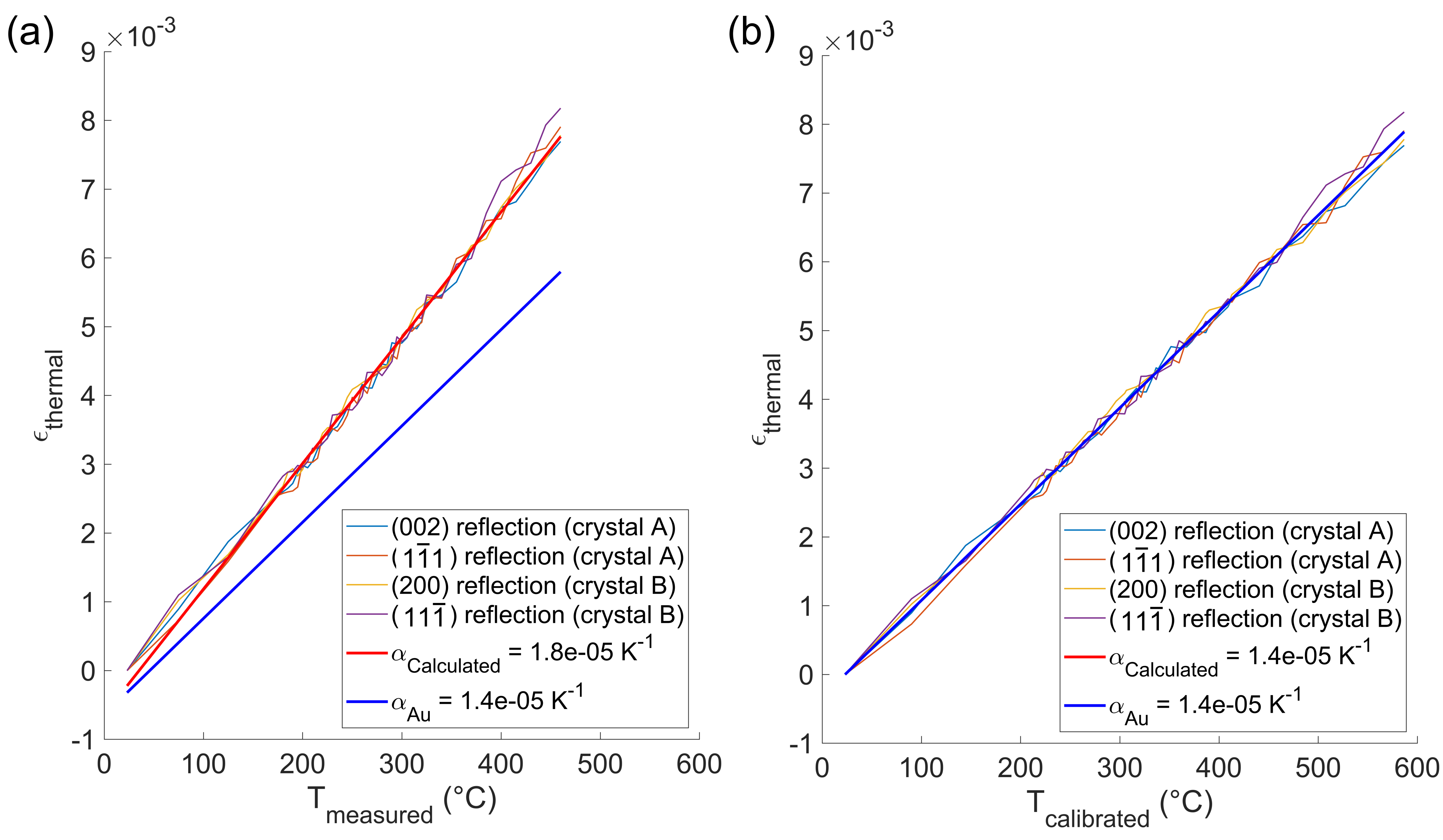} 
    \caption{A comparison of $\epsilon_\mathrm{thermal}$ plotted against (a) $\mathrm{T_{measured}}$ and (b) $\mathrm{T_{calibrated}}$ for each reflection. The $\epsilon_\mathrm{thermal}$ was calculated using Eq. \ref{eq:lattice_constant} and \ref{eq:calibrated_temperature} by tracing the angular position of the detector during annealing. In (a), the measured $\epsilon_{\mathrm{thermal}}$ averaged over all reflections (red line) does not agree with the predicted $\epsilon_{\mathrm{thermal}}$ (blue line) when plotted using $\mathrm{T_{measured}}$. In (b), the measured $\epsilon_{\mathrm{thermal}}$ plotted using $\mathrm{T_{calibrated}}$ (determined by $\alpha_\mathrm{Au}$) overlaps with the the predicted $\epsilon_{\mathrm{thermal}}$ as intended.}
\end{figure}

Fig. \ref{fig:thermal_strain} shows the discrepancy between $\epsilon_\mathrm{thermal}$ plotted against $\mathrm{T_{calibrated}}$ and $\mathrm{T_{measured}}$. The correction for $\mathrm{T_{calibrated}}$ was determined by averaging over the four reflections, with a mean standard deviation of $\sim$8\degree C, indicating good agreement amongst the four reflections.

In this paper all temperatures noted refer to $\mathrm{T_{calibrated}}$ rather than $\mathrm{T_{measured}}$ unless specified. The temperature profile (shown in Fig. \ref{fig:temperature_profile}) has three varying temperature steps so as to capture the entirety of the annealing process whilst retaining a high sampling rate between 209-409\degree C. At low temperatures the migration energy was believed to be too high for noticeable interstitial and vacancy movement, hence larger temperature steps of $\sim$50\degree C were used. Below $\mathrm{T_{H\ddot{u}ttig}}$, the H\"{u}ttig temperature, which marks the threshold for surface atom mobility (127\degree C for Au) \cite{Odarchenko2018}, no changes are expected. Indeed in this experiment, we focus on atomic mobility over tens of nanometers below the surface, as this is the thickness of the layer in which in which Ga implantation causes atomic displacement damage \cite{Hofmann2018a}. Such solid-state diffusion is expected to occur above $\mathrm{T_{Tammann}}$, the Tammann temperature (327\degree C for Au) \cite{Odarchenko2018}, thus from 209\degree C $\sim$5\degree C steps were taken to increase the temperature resolution of our data set in order to capture structural transitions. It is important to note that the H\"{u}ttig and Tammann temperatures are semi-empirical ($\mathrm{T_{H\ddot{u}ttig} = 0.3T_{melting}}$) and ($\mathrm{T_{Tammann} = 0.5T_{melting}}$) and depend on surface texture, orientation, morphology and particle size \cite{Moulijn2001} based on the Gibbs-Thomson equation \cite{Thomson1871}. 
After reaching 409\degree C, the temperature was increased using $\sim$20\degree C steps. This was because much of the FIB damage was expected to already have been removed, based on a study showing the near complete recovery of nanoindented Au microcrystals annealed at $\sim$0.65 times the melting temperature (596\degree C for Au) \cite{Kovalenko2017}.

\subsection{Phasing of the diffraction data} \label{subsection:phasing}
Before the real-space, complex-valued electron density of the crystals can be reconstructed, the phase of the diffracted wavefield must be recovered using an iterative phase retrieval algorithm \cite{Robinson2001}. The use of established phase revival algorithms enables the recovery of the object by iteratively applying real and reciprocal space constraints to an initial guess \cite{Robinson2001,Williams2003,Marchesini2003,Clark2012}. Here, the starting guess for each reflection is of particular interest, as seeding with a good start guess generally results in a faster convergence. At each temperature (except the initial and final temperatures), we could seed the reconstruction with the reconstruction from the previous or succeeding temperature step (see Appendix \ref{appendix:seeding_options}). We instead seed each reconstruction with itself through multiple rounds \cite{Hofmann2020}, with the initial round being seeded with a random guess (see details in Appendix \ref{appendix:phase_retrieval}).

Once the phases were retrieved, the strain fields, $\epsilon\mathbf{(r)}$, projected along the Bragg vector were calculated using:

\begin{equation} \label{eq:strain}
    \epsilon_{hkl}\mathbf{(r)} = \nabla\mathbf{\psi_\mathit{hkl}(r)}\cdot\frac{\mathbf{Q_\mathit{hkl}}}{|\mathbf{Q_\mathit{hkl}}|^2}
\end{equation}

These strain fields are relative to the zero-strain reference of the reconstruction. The assumption of close to zero lattice strain beyond a distance of $\sim$150 nm from the FIB-milled surface has previously been validated \cite{Hofmann2017a,Hofmann2018a}. Thus, we establish a $40 \times 40 \times 40$ voxel volume centred on the bulk ROI's (Fig. \ref{fig:roi}) centre of mass, a region relatively unaffected by FIB-induced strain, as the strain reference for every reconstruction \cite{Phillips2020}. The amplitude threshold is 0.30 for all reconstructions. Furthermore, each final object is eroded by 1 voxel in an attempt to remove a phasing surface artefact caused by steep phase gradients across the air-sample interface convolved with the resolution function \cite{Carnis2019}.

We report the 3D spatial resolution to be $\sim$23 nm. This was determined by differentiating two amplitude line profiles drawn across the air-sample interface and fitting each of them with a Gaussian. The final spatial resolution was the average of the full width at half maximum of the two Gaussians for all reconstructions. We note that the resolution is somewhat dependent on the measurement directions in reciprocal space and the geometry of the sample. Appendix \ref{appendix:spatial_resolution} shows a plot of the spatial resolution of each line profile for each crystal reflection with respect to temperature.

\section{Results and Discussion} \label{section:results_and_discussion}

\subsection{Crystal strain and morphology}\label{subsection:crystal_morphology}
\begin{figure} \label{fig:strain_23_587}
    \centering
    \includegraphics[width=\textwidth,scale=0.5]{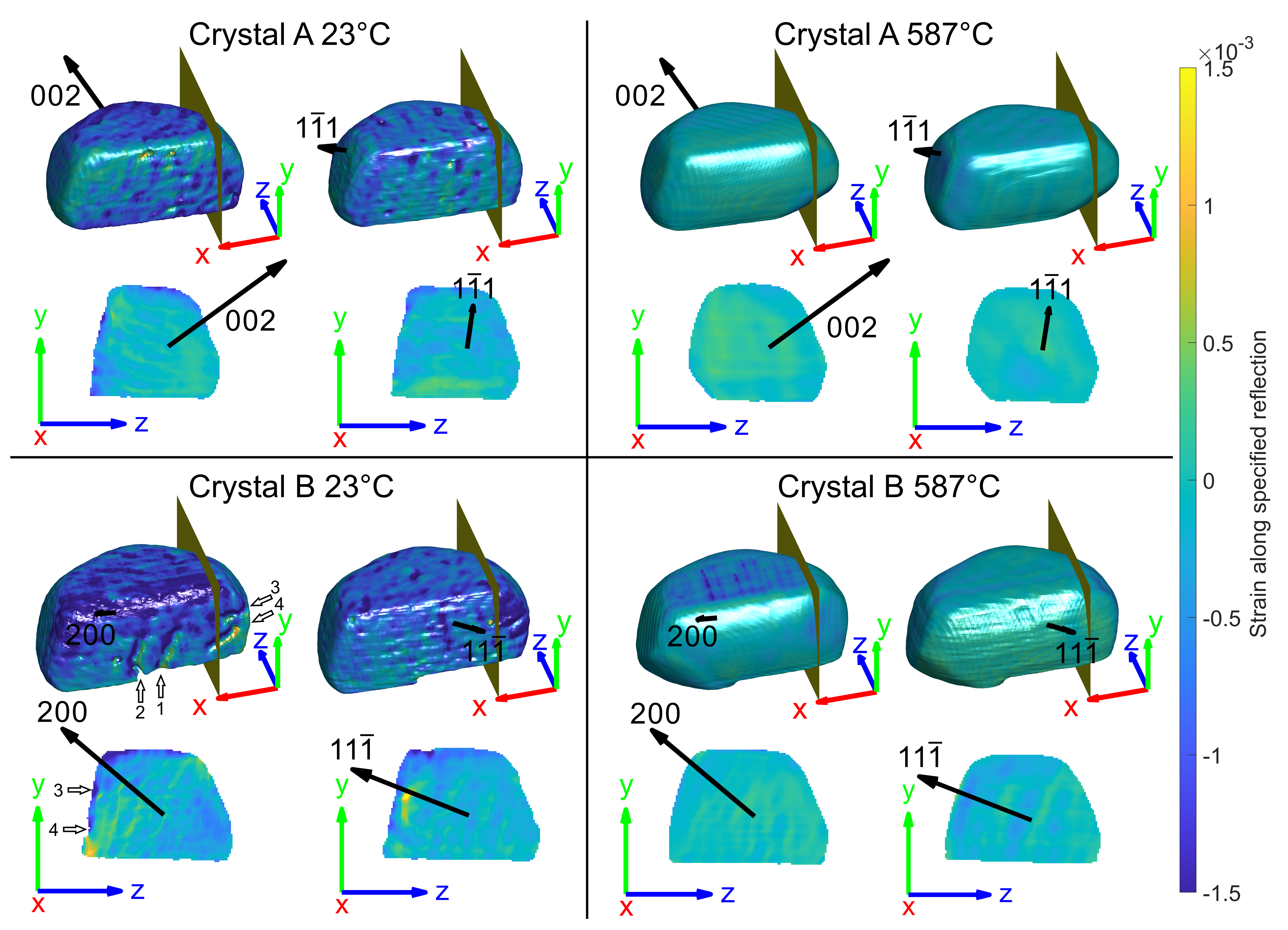}
    \caption{Rendering of the strain in crystal A and B for each reflection at 23\degree C and 587\degree C. In the top row of each quadrant the milled face is shown with a slice perpendicular to the x-axis. The bottom row of each quadrant is the 2D view of the corresponding slice. The small edge at the bottom of crystal B is a twin domain boundary marking a region of missing intensity. The black arrows represent the Bragg vector direction. The coordinate axes are positioned in sample space and plotted with a length of 250 nm. Numbered white arrows highlight trench-like structures that are reminiscent of dislocations. The blue hashed region at 587\degree C on the top face of the $(200)$ reflection for crystal B is a reconstruction artefact (see Section \ref{subsection:phasing}) and does not physically represent a compressive layer.}
\end{figure}

Fig. \ref{fig:strain_23_587} shows a comparison of the strain found along each reflection, $\mathbf{Q_\mathit{hkl}}$, of the FIB-milled crystals between the initial and final temperature.
At room temperature, the reconstructed crystal morphologies are rough, especially on the surface of the milled face where complex sputtering and redeposition processes compete, which results in a pitted surface morphology. In Fig. \ref{fig:strain_23_587} the four trench-like features visible on the surface of the $(200)$ reflection for crystal B are near-surface dislocations caused by the milling process. These manifest as pipes of missing intensity in the amplitude of the reconstruction and are at the centre of a phase vortex \cite{Clark2015,Ulvestad2015a,Hofmann2017a,Dupraz2017,Hofmann2020}. The phase vortices of the $(200)$ reflection are identified in Appendix \ref{appendix:200_phase}. Dislocations are visible only when $\mathbf{Q_\mathit{hkl}}\mathbf{\cdot b}\neq0$ \cite{Williams2009} and therefore are not necessarily visible for all reflections. For crystal B, the dislocations are visible only for the $(200)$ data. From two reflections alone it is not possible to uniquely identify the Burgers vector of these dislocations, although it is likely that it is of $(a/3)<110>$ type \cite{Hofmann2017a}. 

The room temperature 2D strain slices for each reflection are comparable in magnitude, with a maximum strain of $\sim$1.5$\times 10^{-3}$, to published data for glancing incidence FIB damage in Au \cite{Hofmann2018a}. The same publication reports a thin compressive region followed by a tensile region along the milled face, which is prominently observed here in the $(200)$ reflection for crystal B. A previous normal incidence FIB-milling study on Au microcrystals at room temperature \cite{Hofmann2017a} reported a compressive layer at the top of the samples, caused by normal incident angle ions impacting the top of the crystal from the tails of the FIB probe. This is seen in the $(002)$ reflection for crystal A and both the $(200)$ and $(11\bar{1})$ reflections for crystal B.

After heating to 587\degree C, the strain and the crystal morphology have changed dramatically (Fig. \ref{fig:strain_23_587}) and presents a relatively constant strain in the 2D slices. Additionally, the surface morphology is observed to become smooth and evolution of the facets is observed. Supplementary videos (SV) SV1 and SV2 show the evolution of the surface morphology and internal strain for each crystal. In Fig. \ref{fig:strain_23_587} the dislocations or pits on the milled face in the $(200)$ reflection disappear after roughly (1) 296\degree C, (2) 352\degree C, (3) 372\degree C and (4) 380\degree C, indicated by the elimination of phase vortices, as surface restructuring is well underway. These temperatures are mostly consistent with being above the Tammann temperature (327\degree C) for Au.

\subsection{Facet area}\label{subsection:facet_area_plots}
\begin{figure} \label{fig:facet_directions}
    \centering
    \includegraphics[scale=0.7]{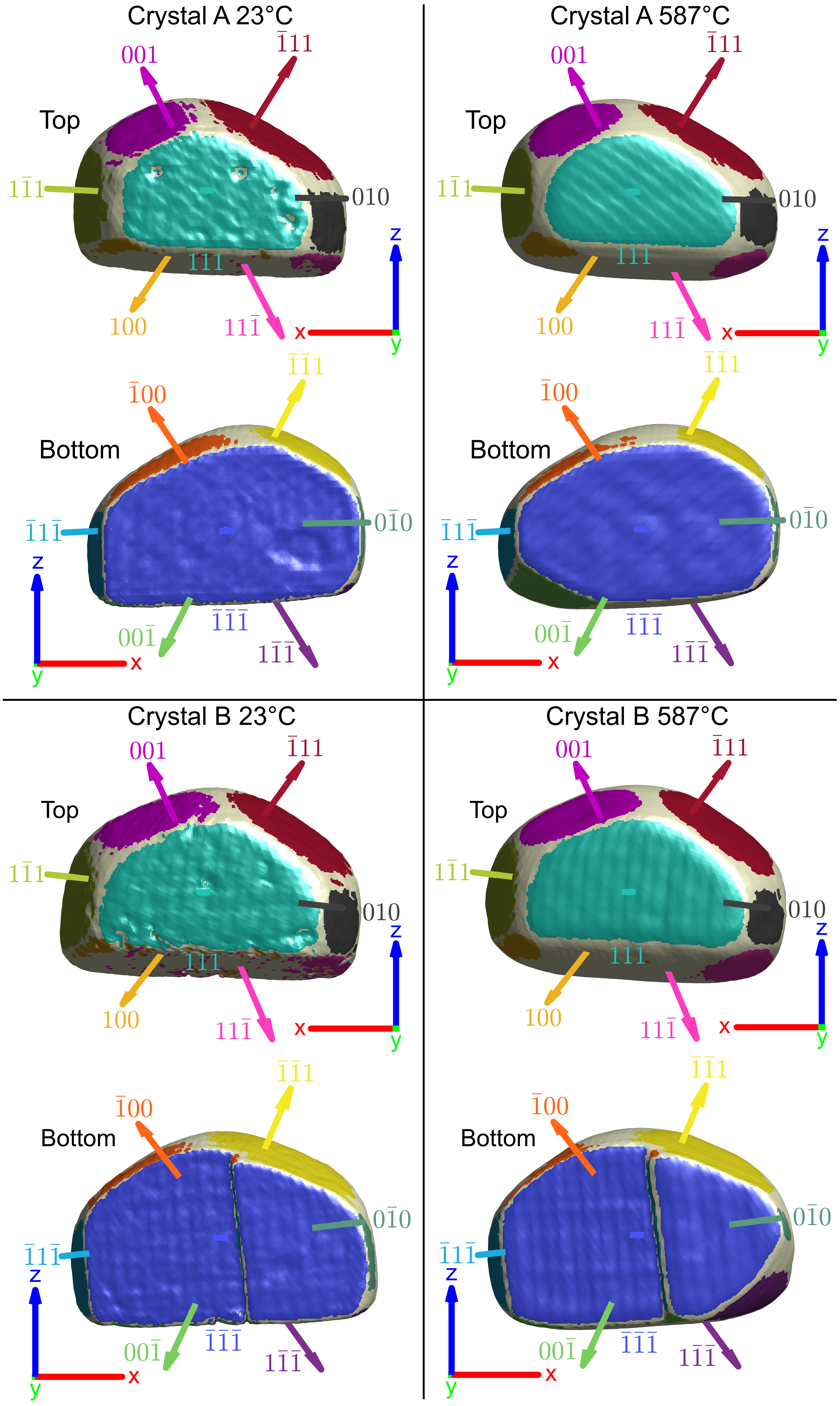}
    \caption{Normal vectors for each separately coloured facet are shown at $23 \mathrm{\degree C}$ and $587 \mathrm{\degree C}$ for both crystals. Note only vectors pointing out of the page in the top and bottom views are drawn for clarity. The coordinate axes are positioned in sample space and plotted with a length of 250 nm.}
\end{figure}

With increasing temperature, both milled faces expanded outwards along the z-direction and became rounder at the bottom, demonstrating the formation of low-index facets, shown in Fig. \ref{fig:facet_directions}, as a means to reduce the overall free surface energy. Facet areas were computed by calculating the area of the crystal whose isonormal is within a $19\degree$ angular threshold of a given facet normal unit vector. In principle a maximum angular threshold of $27.4\degree$ could be used, which is half the angle between \{111\} and \{100\} planes. Our chosen tolerance is below this maximum value since not all facets are fully developed, and the edges of the crystal are rounded and thus should not be classified as part of a facet.

\begin{figure} \label{fig:facet_sizes}
    \centering
    \includegraphics[scale=0.67]{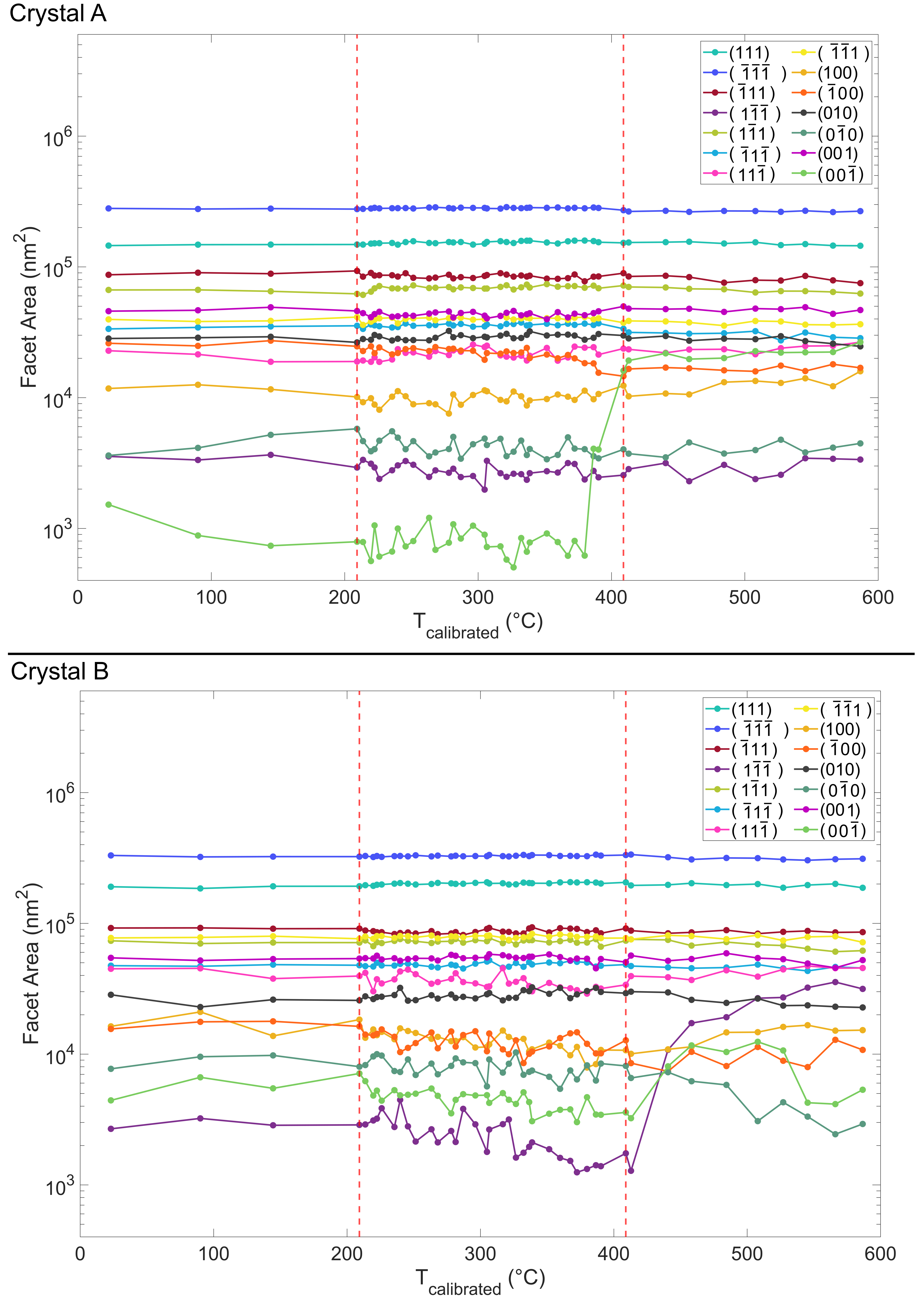}
    \caption{Low-index facet areas of crystal A and B as a function of temperature. See Fig. \ref{fig:facet_directions} for facet locations on the crystal.}
\end{figure}

Fig. \ref{fig:facet_sizes} and supplementary video SV3 show facet area evolution as temperature increases for both crystals. Fig. \ref{fig:strain_23_587} shows that the facet at the bottom right edge of crystal A that appears after 410\degree C (see supplementary video SV1) is $(00\bar{1})$ and the facet at the bottom left edge of crystal B that appears after 441\degree C (see supplementary video SV2) is $(1\bar{1}\bar{1})$. The identification of these facets is based on Fig. \ref{fig:facet_directions} and supplementary video SV3. The formation of low-index facets has been suggested to relieve strain \cite{Kim2018}, and contribute to thermodynamic equilibrium \cite{Kovalenko2015}. However, we do not fully reach the Winterbottom thermodynamic equilibrium shape \cite{Winterbottom1967} due to finite measurement time. For Au, the order of free surface energies for low index facets is $\{111\} < \{100\} < \{110\}$ based on ab initio techniques that concur with the number of nearest-neighbour broken bonds \cite{Galanakis2002}. Thermodynamically, the formation of a $(1\bar{1}\bar{1})$ facet is most favourable, shown in crystal B. However, the production of the $(00\bar{1})$ facet in crystal A and the disappearance of the $(00\bar{1})$ facet in crystal B after 527\degree C are not improbable as normal movement of facets in dislocation‐free single crystals can be restricted by the magnitude of the nucleation barriers if the crystal is larger than a few nanometers \cite{Mullins2000}. In crystal B, the $(00\bar{1})$ facet shrinks after 527\degree C but the $(1\bar{1}\bar{1})$ facet grows, suggesting that the latter is a lower energy configuration.

\subsection{Normalised displacement field cross-correlation matrix}\label{subsection:XC_disp}
To pinpoint transitions in strain, we consider a Pearson $r$ correlation matrix of the displacement fields between different temperatures for each reflection. When analysing single reflections, this method has been used \cite{Yau2017a,Ulvestad2015c} to infer structural transitions. The value of $r$ between each combination of temperatures was computed using Eq. \ref{eq:XC} \cite{Ulvestad2015c}: 

\begin{equation} \label{eq:XC}
     r(x,y) = \frac{\sum\limits_{n}(x_n-\bar{x})(y_n-\bar{y})}{\sqrt{\sum\limits_{n}(x_n-\bar{x})^2}\sqrt{\sum\limits_{n}(y_n-\bar{y})^2}}
\end{equation}

where $x_n$ is the value for a given voxel and temperature, $\bar{x}$ is the mean of the entire array for that temperature, $y_n$ is the value for the same voxel at another temperature and $\bar{y}$ is the mean of the entire array at that temperature. By definition, $r = 1$ for the diagonal because the same array is perfectly correlated to itself. Before computing $r$, each crystal was shifted such that its  was in the middle of the $256 \times 256 \times 256$ voxel array, which was subsequently cropped to a $98 \times 186 \times 126$ voxel array, still fully encapsulating the crystal, to speed up $r$ computation. 

\begin{figure} \label{fig:XC_displacement}
    \centering
    \includegraphics[width=\textwidth,scale=0.5]{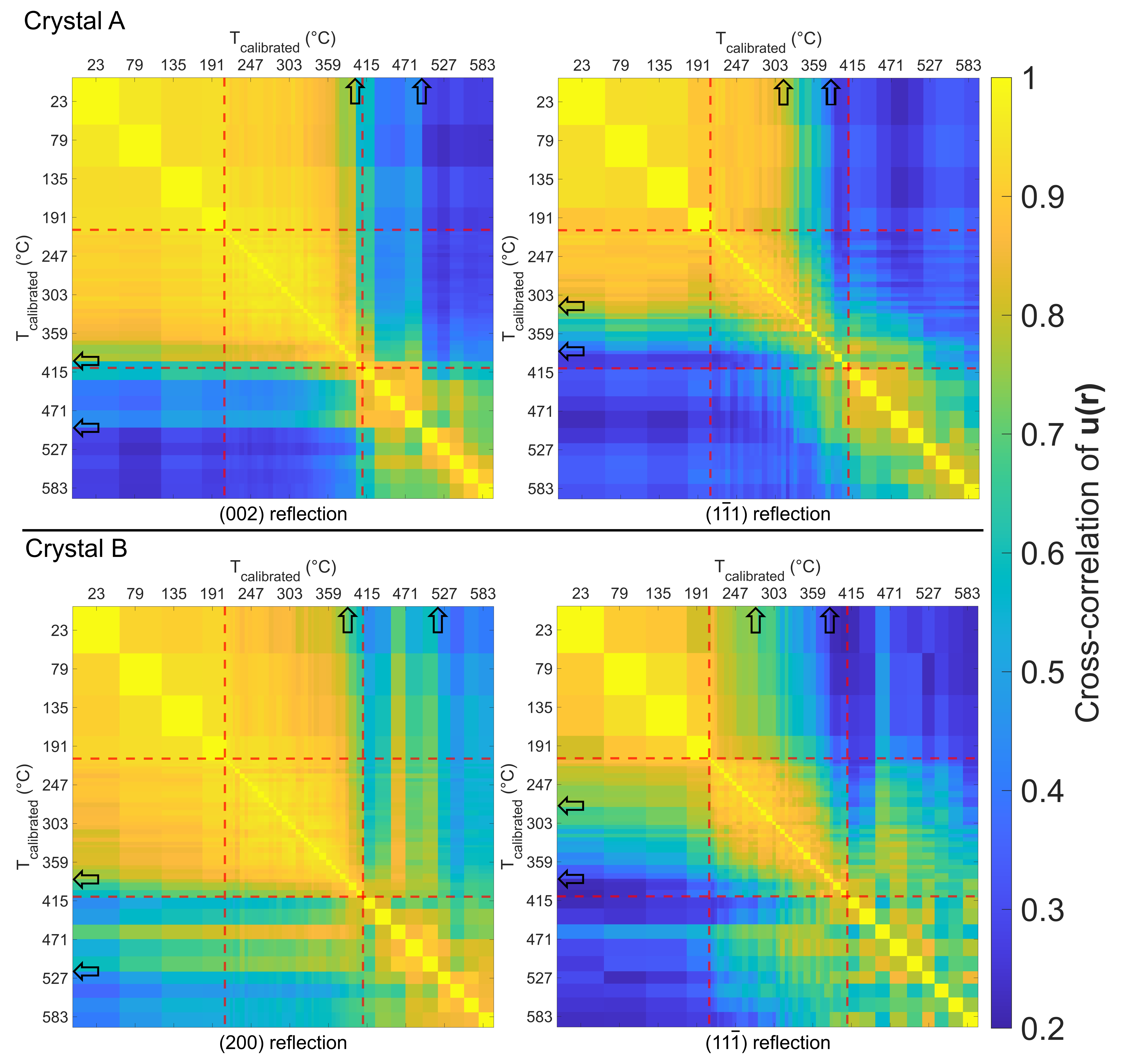}
    \caption{Reflection-specific $r$ matrices between $\mathbf{u_{\textit{hkl}}(r)}$ for different temperatures. The difference in $r$ values corresponds to structural evolution. The $(002)$ reflection for crystal A shows an abrupt $r$ change after 410\degree C, suggesting the abrupt removal of most defects. This is less evident in the $(11\bar{1})$ reflection for crystal B, which shows many subtle structural changes as a function of temperature. The red dashed lines indicate where the temperature sampling changed. Black arrows represent the most noticeable changes in displacement.}
\end{figure}

The $r$ matrix is shown in Fig. \ref{fig:XC_displacement} for $\mathbf{u_{\textit{hkl}}(r)}$. Before using Eq. \ref{eq:phase} to calculate $\mathbf{u_{\textit{hkl}}(r)}$, phase ramps were removed by re-centring the Fourier transform of the complex electron density and a phase unwrapping algorithm \cite{Cusack2002} was used to remove phase wraps. Furthermore, we accounted for arbitrary phase offsets by establishing a $40 \times 40 \times 40$ voxel volume centred about the bulk ROI's centre of mass as the zero-phase reference for every reconstruction
, similar to the procedure adopted for strain in Section \ref{subsection:phasing}. 

Structural transitions in the sample are identified by sharp changes in $r$ presented in Fig. \ref{fig:XC_displacement}. Surprisingly the two reflections from crystal A have different temperatures where the most significant changes occur: abruptly after 410\degree C in the $(002)$ reflection and less abruptly at 380\degree C in the $(1\bar{1}1)$ reflection. These transitions are also reflected in the facet area plot for crystal A in Fig. \ref{fig:facet_sizes}. The discrepancy between transition points could be due to the $(00\bar{1})$ facet forming at 380\degree C but grows significantly larger after 410\degree C.

Some of these abrupt changes in $r$ occur as a result of sudden relaxation events associated with the removal of a defect, such as dislocations (3), removed at 372\degree C, and (4), removed at 380\degree C, from Fig. \ref{fig:strain_23_587} and supplementary video SV2. In crystal A the $(002)$ reflection has reached a steady annealed state after 410\degree C, but this annealed state is less defined in the other reflections. Examining the same family of reflections can lead to different conclusions, as crystal B shows less abrupt transitions within the $r$ matrices (especially the $(200)$ reflection) that suggest a more gradual recovery of the lattice. In addition to these prominent transitions within the $r$ matrix, less pronounced fluctuations are present and here are attributed to the oscillation between intermediate annealing states.

The differences in observations between reflections from the same crystal arises because not all defects are visible in every reflection and the projection of associated displacement fields varies for each combination of Burgers vector and scattering vector. This variance in material transformation points highlights the importance of multi-reflection analysis, as it can be misleading to derive conclusions from just a single $r$ matrix.

\subsection{Normalised morphology cross-correlation matrix}\label{subsection:XC_mask}
\begin{figure} \label{fig:XC_mask}
    \centering
    \includegraphics[width=\textwidth,scale=0.5]{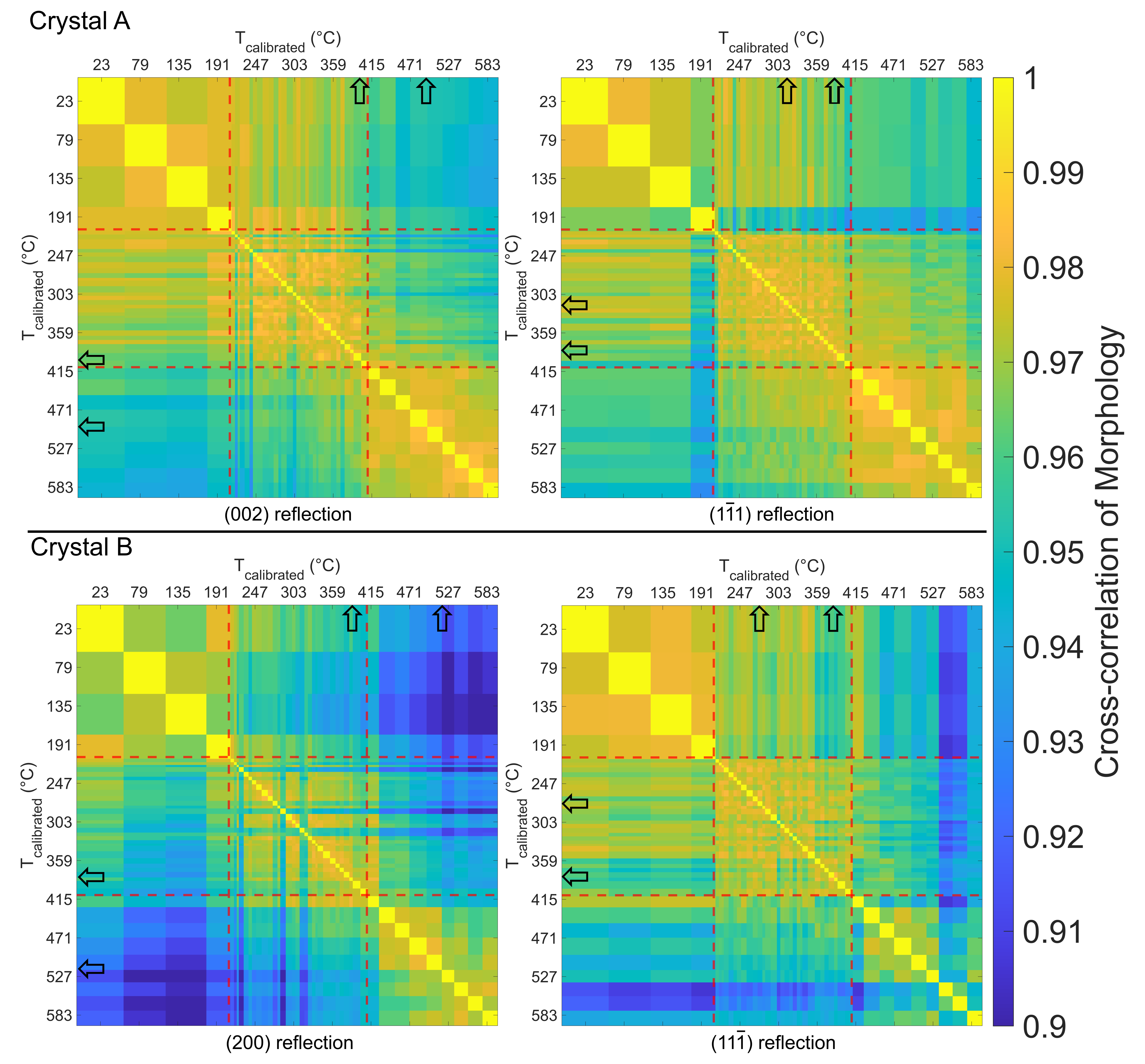}
    \caption{Reflection-specific $r$ matrices for the morphology at different temperatures. The difference in $r$ values corresponds to morphology evolution. There is a more gradual $r$ transition as the temperature increases compared to Fig. \ref{fig:XC_displacement}. The red dashed lines indicate where the temperature sampling changed. Black arrows represent the most noticeable changes in displacement as observed in Fig. \ref{fig:XC_displacement}.}
\end{figure}

To separate morphology-related changes from lattice displacement changes, the $r$ correlation matrix of the morphology is shown in Fig. \ref{fig:XC_mask}. The morphology of a  crystal was created by applying a 0.30 threshold to the reconstructed amplitude.

There is good agreement between the morphology $r$ matrices and the facet area plots. For instance, the disappearance of the $(00\bar{1})$ facet in crystal B shown in Fig. \ref{fig:facet_sizes} after 527\degree C can also be clearly identified in the $(11\bar{1})$ reflection in Fig. \ref{fig:XC_mask}. Notable changes in $r$ for $\mathbf{u_{\textit{hkl}}(r)}$ in Fig. \ref{fig:XC_displacement}, marked by hollow black arrows in Fig. \ref{fig:XC_displacement} and Fig. \ref{fig:XC_mask} are different to changes in $r$ for the morphologies shown in Fig. \ref{fig:XC_mask}. There are less abrupt $r$ changes in Fig. \ref{fig:XC_mask}, suggesting that annealing causes smaller incremental changes in morphology, different from the sudden changes in $\mathbf{u_{\textit{hkl}}(r)}$ (see Fig. \ref{fig:XC_displacement}). However, there are changes in $\mathbf{u_{\textit{hkl}}(r)}$ that cannot be seen in the facet area plots in Fig. \ref{fig:facet_sizes} or the morphology matrices in Fig. \ref{fig:XC_mask}, namely around $\sim$310\degree C in the $(1\bar{1}1)$ reflection in Fig. \ref{fig:XC_displacement}. Changes in strain that cannot be explained by changes in morphology can be examined directly in the next section.

\subsection{Average lattice strain in regions of interest}\label{subsection:strain_in_roi}
\begin{figure} \label{fig:roi}
    \centering
    \includegraphics[width=\textwidth]{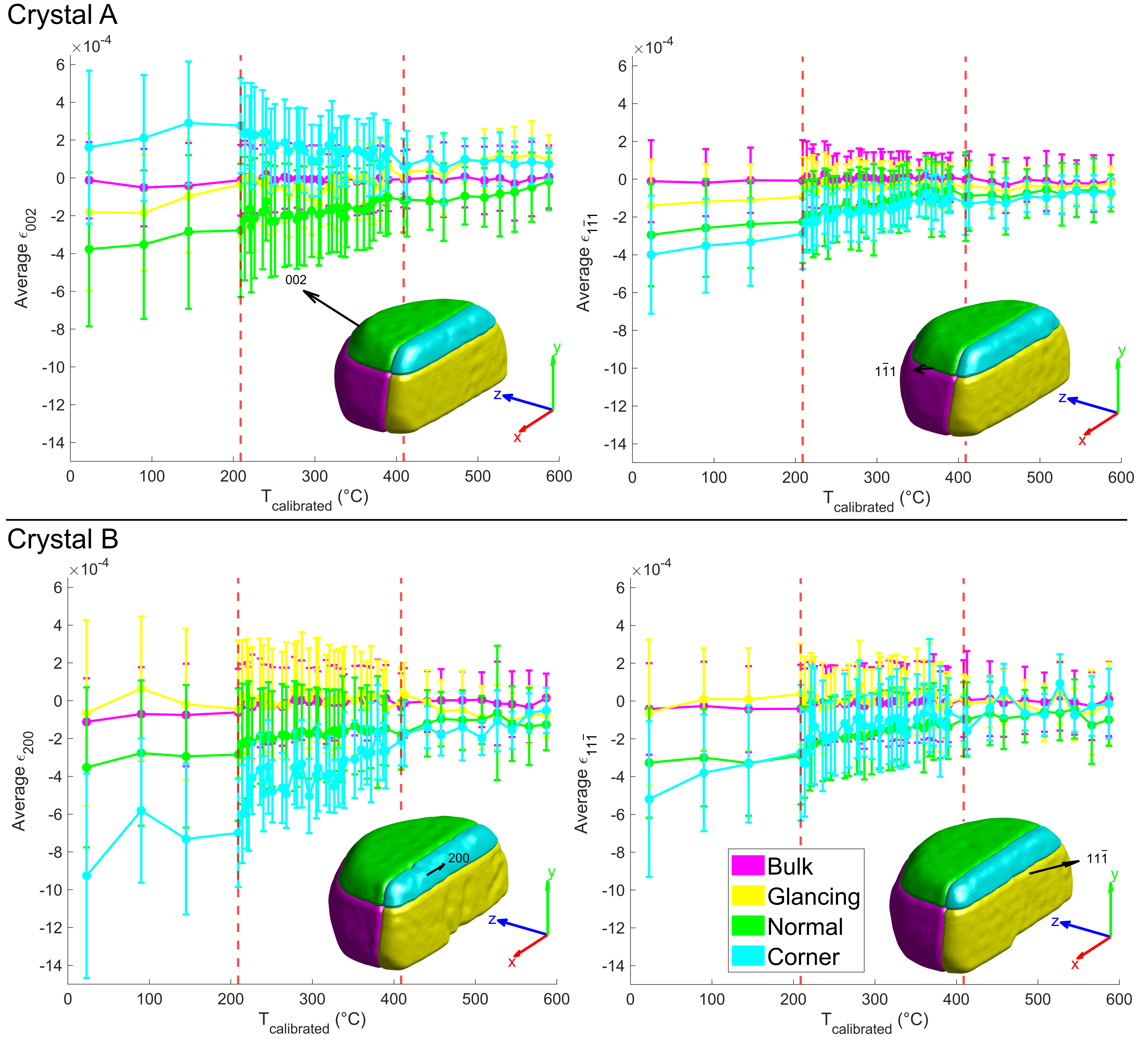}
    \caption{Average lattice strain in different ROIs as a function of temperature. Vertical bars correspond to the standard deviation of the strain. Over the annealing cycle, the average strain in each reflection gradually decreases and stagnates after $\sim$410\degree C, as does the magnitude of the standard deviation. The magnitude and direction of the average strain is reflection-dependent, especially in the corner ROIs. The coordinate axes are positioned in sample space and plotted with a length of 250 nm. The red dashed lines indicate where the temperature sampling changed. The rendered crystals show the locations of the ROIs.}
\end{figure}

We will now consider the average strain within particular regions of the crystal, illustrating key differences in the local lattice environment. The average lattice strain for a reflection was computed by calculating the strain for each voxel using Eq. \ref{eq:strain}, and taking the mean over all voxels with an electron density magnitude greater than 0.30. Each crystal has been divided into four ROIs, shown in Fig. \ref{fig:roi}, according to how the Ga-beam has interacted with the sample. The ROIs are labelled glancing, normal, corner and bulk. The glancing ROI is defined as 100 nm (20 voxels) from the surface of the milled face where the Ga beam impinged on the sample at glancing incidence. Although the FIB milling was intended to be normal to the substrate, the milled face is $\sim$80\degree\ from the local sample surface normal. This is caused by the tail of the Gaussian-like FIB probe profile not being able to fully reach the bottom of the crystal. The normal ROI is 100 nm from the top of the crystal, corresponding to the region where the FIB beam arrives approximately normal to the sample surface. The corner ROI is the volume that joins the glancing and normal regions. Previously the corner was seen to have the most defects \cite{Hofmann2018a}. The bulk ROI is the remaining part of the crystal. The average lattice strain for the different ROIs at every temperature point is shown in Fig. \ref{fig:roi}.


In Fig. \ref{fig:roi}, the strain within the bulk ROI remained approximately zero for all reflections and temperatures because no FIB-induced strain was present prior to annealing. The room temperature data is similar to the values reported previously \cite{Hofmann2018a}. As the temperature increases, we see that FIB damage is removed, and the average strain converges to zero.  We observe that the glancing and normal ROIs behave similarly regardless of reflection, with the normal ROI having more compressive strain than the glancing ROI. This suggests that the FIB tail imparts more damage to the top face of the crystal than does glancing incidence milling to the milled face, in agreement with previous work \cite{Hofmann2018a,Hofmann2017a}. The most reflection-dependent ROI is the corner, where both the average magnitude and the sign of strain vary. For the corner ROI in crystal A, the $(002)$ reflection shows tensile strain whereas the $(1\bar{1}1)$ reflection shows compressive strain of similar average magnitude for the same crystal. The average strain magnitude can be dependent on the reflection. This is observed in the corner ROI for crystal B at room temperature, where the $(200)$ reflection has nearly twice the average strain magnitude of the $(11\bar{1})$ reflection. These significant differences would not be evident from single reflection BCDI experiments.

The average strain curves approach zero at $\sim$410\degree C in Fig. \ref{fig:roi}. This is above the Tammann temperature (327\degree C) and matches the temperature where facet change is observed for crystal A in Fig. \ref{fig:facet_sizes} and where structural changes occur for the $(002)$ reflection for crystal A in Fig. \ref{fig:XC_displacement}. Above  $\sim$410\degree C there is little strain evolution, indicating that most of the damage in the crystal has been annealed out. Further evidence for this transition point can be seen in the standard deviation magnitudes of Fig. \ref{fig:roi}. Below $\sim$410\degree C, the standard deviation magnitudes for the glancing, normal and corner ROIs (plotted in Appendix \ref{appendix:std_avg_strain}) are greater than that of the bulk ROI. Above $\sim$410\degree C, the standard deviations across all ROIs are similar, indicating that the average strain across all ROIs has become equally homogeneous. The removal of strain inside the crystals can also be seen by comparing the maximum intensity slices of the CXDPs in Appendix \ref{appendix:CXDP_comparison} and noting the change in fringe sharpness before and after annealing. The evolution of these CXDP slices can be seen in supplementary videos SV4 and SV5.

We also note that the differences in the rate of average strain removal seem to depend on the severity of the initial FIB damage. It appears that having a higher initial average strain magnitude leads to a greater initial change in average strain. This is most apparent when comparing the rate of average strain change in the corner, glancing and normal ROIs for any reflection, as well as solely the corner ROIs in crystal B. 

\subsection{Analysis technique comparisons}\label{subsection:Analysis_technique_comparisons}
It is interesting to compare information extracted from the facet area plots, $r$ matrices and average strain plots. While we see that the facet area plots and displacement $r$ matrices can highlight distinct transition points, only major changes can be confidently identified. In particular, crystal transitions can be readily identified in the facet area plots and displacement $r$ matrices but remain harder to resolve when observing average strain plots and morphology $r$ matrices. The primary advantage of the average strain plots over the consideration of $r$ matrices is the ability to identify the magnitude and the direction of strains in various ROIs, which can confirm whether the FIB-induced strain has indeed been removed.

The seemingly abrupt structural evolution in the facet area plots and displacement $r$ matrices contrasts the more uniform rate of strain change observed in the average strain plots and morphology $r$ matrices. The facet area plots and displacement $r$ matrices portray annealing as a process controlled by sudden substantial changes, implying a rapid evolution in morphology and removal of strain after a certain temperature is reached, as shown by the formation of facets in Fig \ref{fig:facet_sizes} and the $r$ for $\mathbf{u_{\textit{hkl}}(r)}$ in Fig \ref{fig:XC_displacement}. For the same reflection in the morphology $r$ matrices in Fig \ref{fig:XC_mask} and in the average strain plots in Fig. \ref{fig:roi}, annealing appears as a gradual process, by which the change in temperature leads to a progressive change in lattice relaxation and facet edge softening. The trench-like dislocations observed in Fig.  \ref{fig:strain_23_587} gradually decrease in size but also fully disappear once certain temperatures are reached, seen in supplementary video SV2, exhibiting both gradual and sudden changes in crystal structure as a result of annealing.

The accuracy of these observations is limited by our spatial resolution and sampling time. To better resolve defect dynamics, faster scans and higher photon flux are required, which the Advanced Photon Source Upgrade Project promises to deliver. 
Regardless, the conclusions drawn from each of the analysis techniques are complementary and reveal different aspects of the crystal behaviour, but point towards an annealing temperature of 380-410\degree C. These methods are suitable for the analysis of individual reflections, but can benefit from the incorporation of multiple reflections, enabling more comprehensive comparisons to be made.

\subsection{Diffusion considerations}\label{subsection:Diffusion_considerations}
Scan times on the order of minutes for BCDI rocking curves limit fast structural changes from being resolved. However, we are able to capture structural evolution on the timescale of the rocking scans, such as the formation of facets, disappearance of dislocations and relative changes in $\mathbf{u_{\textit{hkl}}(r)}$. These changes can be interpreted in terms of the diffusion of Ga in Au, the self-diffusion of Au at the surface or within the bulk of the crystal and the escape of dislocations to the free surface. 

\begin{figure} \label{fig:diffusion_plot}
    \centering
    \includegraphics[width=\textwidth]{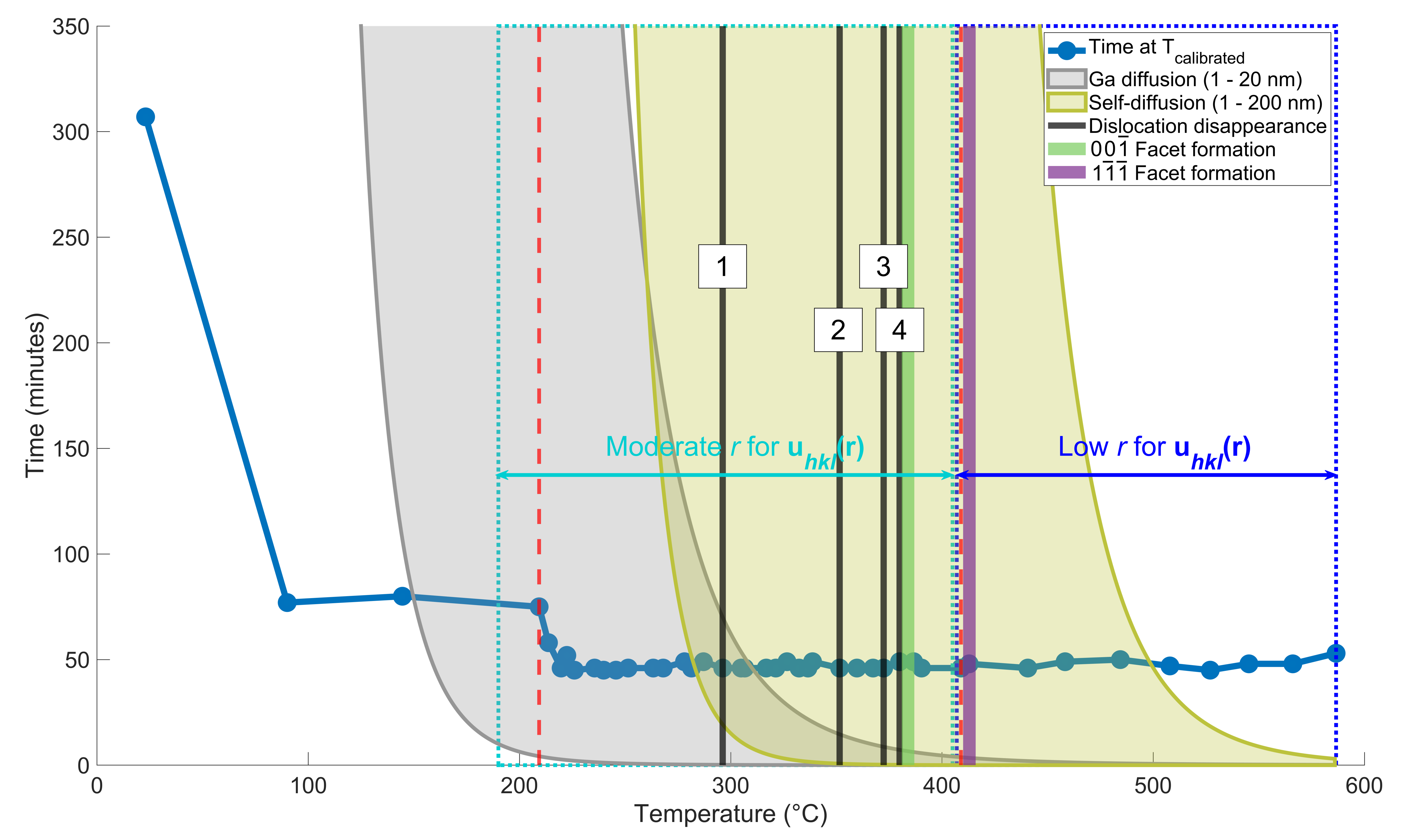}
    \caption{The time required for the bulk diffusion of Ga over lengths from 1 to 20 nm (light grey region) and Au diffusion over distances from 1 to 200 nm (dark yellow region) throughout the annealing. Superimposed are temperature points where dislocations in Fig. \ref{fig:strain_23_587} disappeared, temperature points where facets have formed in Fig. \ref{fig:facet_sizes}, and temperature regions where the $\mathbf{u_{\textit{hkl}}(r)}$ $r$ values changed noticeably (see Fig. \ref{fig:XC_displacement}). The dotted blue line shows the time-temperature trajectory of the Au sample in this study, indicating the time spent at each annealing temperature. The red dashed lines indicate where the temperature sampling changed.}
\end{figure}

Fig. \ref{fig:diffusion_plot} shows the temperature and time required for diffusion of Ga over lengths from 1 nm to 20 nm, as well as self-diffusion of Au over 1 nm to 100 nm. We use diffusion coefficients for these two systems as they provide an indication of relative atomic mobility at various temperatures. The diffusion coefficient of Ga in Au was obtained from \cite{Gupta1987}, while the self-diffusion of Au was obtained from \cite{Makin1957}. We determine the time required for 1 nm of diffusion as a function of temperature, for each scenario, as shown by the exponential curves in Fig. \ref{fig:diffusion_plot}. The penetration depth of Ga under the FIB conditions used is $\sim$20 nm \cite{Hofmann2018a}, thus we calculate the time required for a diffusion length of 20 nm as the upper limit of Ga diffusion. Diffusion lengths between 1 and 20 nm are shown as a grey region between the 1 nm and 20 nm curves. Similarly, for the self-diffusion of Au, diffusion lengths between 1 and 200 nm are relevant, as the greatest change in facet area is about $4000\mathrm{\ nm^2}$ (Fig. \ref{fig:facet_sizes}). Superimposed on this time-temperature plot is the temperature history of the present Au samples (dotted blue line), indicating the time spent at each temperature during the annealing process.

To Fig. \ref{fig:diffusion_plot} we can further add the changes in $\mathbf{u_{\textit{hkl}}(r)}$ cross-correlation observed as a function of temperature (Fig. \ref{fig:XC_displacement}). From $\sim$190\degree C until $\sim$410\degree C there is a gradual reduction in the cross-correlation value. Considering the diffusion lengths of Ga and Au, this change at lower temperatures up to $\sim$280\degree C is likely to be driven by bulk Ga diffusion and Au surface diffusion. Above this temperature bulk Au self-diffusion may be responsible. At $\sim$410\degree C Fig. \ref{fig:XC_displacement} shows a rapid change in cross correlation coefficient, suggesting major structural changes as discussed above. The diffusion curves in Fig. \ref{fig:diffusion_plot} suggest that this is associated with the rearrangement of Au rather than Ga atoms. 

On Fig. \ref{fig:diffusion_plot} we have superimposed the temperatures at which removal of dislocations is observed (black lines labelled 1 to 4, consistent with dislocation numbers in Fig. \ref{fig:strain_23_587}), a well as the temperatures at which the $(00\bar{1})$ and $(0\bar{1}\bar{1})$ facets are formed (green and purple lines respectively, consistent with Fig. \ref{fig:facet_directions} and \ref{fig:facet_sizes}). Apart from the removal of dislocation 1, which is at the boundary between Ga diffusion and Au diffusion dominated areas, all other structural changes appear to be driven by Au migration.

\section{Conclusion} \label{section:conclusion}
We have presented an \textit{in situ} BCDI study of the annealing of FIB-induced damage in Au microcrystals. Two separate crystals with two reflections each were measured, confidently switching between reflections and samples, during an annealing cycle. By tracking the change in detector position, we could use the thermal strain of the microcrystals for temperature calibration. We have also presented a new approach that monitors the reduction of free surface energy by following morphology evolution through the formation of facets. The annealing process was analysed using facet areas, cross-correlation matrices and average strain plots, which provide complementary information. Surprisingly, two different reflections from the same crystal can show different behaviour. This highlights the need to measure multiple reflections in order to gain a complete picture of sample evolution. Full annealing was achieved at a temperature of 380-410\degree C. However, the complexity of the process becomes apparent when trying to evaluate the rate of strain removal, as the facet area plots and displacement $r$ matrices portray annealing as an abrupt process, while the morphology $r$ matrices and average strain plots show much more gradual changes. The gentle evolution of the crystal below $\sim$280\degree C is likely the result of bulk Ga diffusion and Au surface diffusion whilst more substantial changes (i.e. removal of dislocations and formation of facets) above $\sim$280\degree C are likely dominated by the self-diffusion of Au. This experiment demonstrates that annealing is a viable option for the removal of FIB damage as evident by a decrease in strain heterogeneity inside the crystal, removal of defects on the crystal surface and reduction of mean strain in FIB-damaged areas. Annealing can thus be used as a method to remove FIB-induced sample damage prior to further modification of a sample e.g. nanoindentation, bending etc. 

We anticipate that the techniques developed here could be used to pioneer novel \textit{in situ} BCDI measurements in complex environments for multiple samples and reflections simultaneously. The processed diffraction patterns, final reconstructions and data analysis scripts are publicly available at: https://www.doi.org/10.5281/zenodo.4423216. 

\pagebreak

\appendix

\section{34-ID-C heater cell} \label{appendix:heater}
\begin{figure} \label{fig:heater}
    \centering
    \includegraphics[width=\textwidth,scale=0.5]{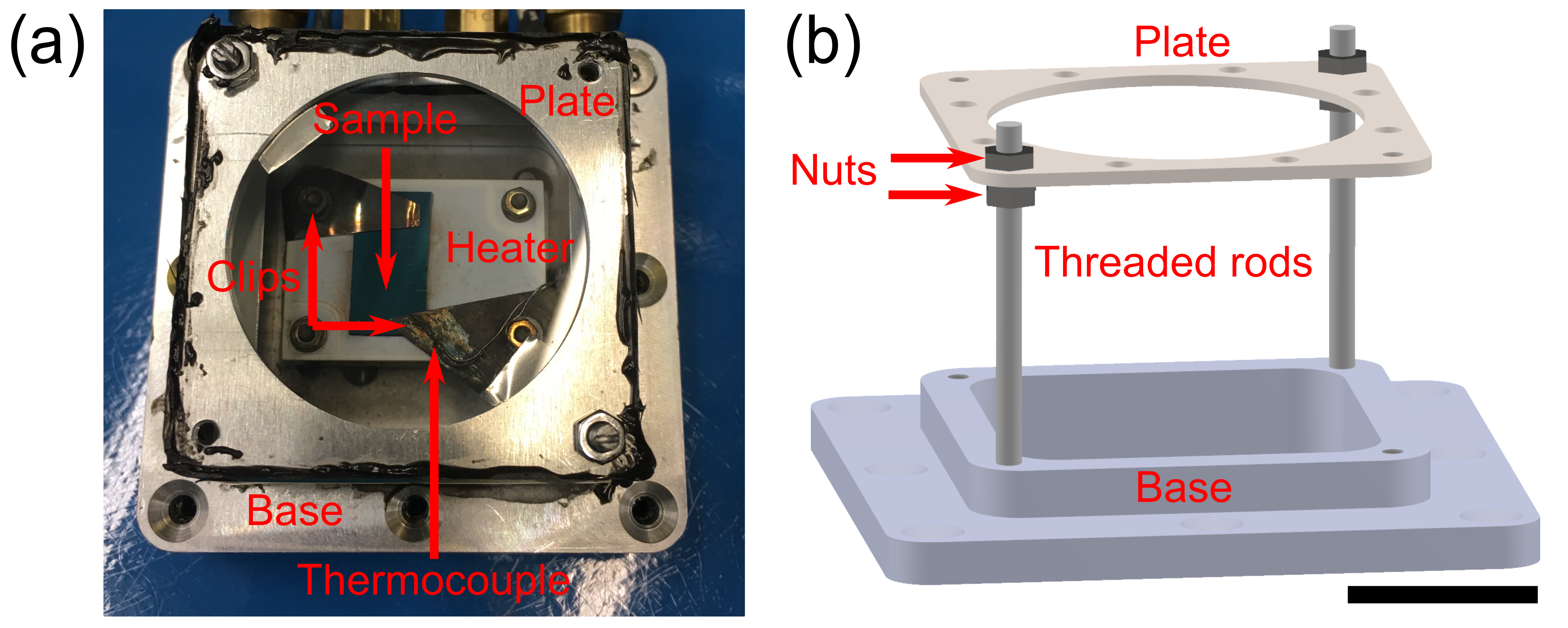} 
    \caption{(a) The sample clipped in the heater cell as seen through the heater enclosure window. Note the thermocouple is attached to the clip, which contacts the silicon substrate. This results in thermal resistance between the Au crystals and the thermocouple, causing the discrepancy between the measured and actual temperature of the sample. (b) A rendering of the furnace enclosure cover. The scale bar corresponds to 20 mm.}
\end{figure}

Fig. \ref{fig:heater}(a) shows the heater within a furnace enclosure. Fig. \ref{fig:heater}(b) shows a rendering of the enclosure cover, consisting of a stainless-steel cover plate attached to a base using stainless-steel threaded rods and nuts. Only two rods were used to minimise the obstruction of incident and reflected X-rays. The circular hole in the plate was covered with a quartz window to allow observation of the sample with a confocal microscope with a $5\times$ objective lens during the annealing experiment. The sides of the furnace enclosure cover were covered with metalized Kapton (not shown) to allow for transmission of X-rays  and favorable thermal transport for cooling of the film.


\section{Effect of seeding on the retrieved electron density} \label{appendix:seeding_options}
This section addresses how the seeding of the phase retrieval algorithm influences the final object's amplitude and strain. For a CXDP at an arbitrary temperature, we consider the following options:

\begin{enumerate}
    \item Seeding with the reconstruction from the temperature step immediately below the current temperature, except if the current temperature is 23\degree C, in which case a random guess is used.
    \item Seeding with the reconstruction from the temperature step immediately above the current temperature, except if the current temperature is 587\degree C, in which case a random guess is used.
    \item Seeding with a random guess regardless of the current temperature.
\end{enumerate}

A simpler phase retrieval algorithm compared to the one listed in Appendix \ref{appendix:phase_retrieval} was implemented to exaggerate the differences in the final reconstructions. A pattern of 20 error reduction (ER) and 180 hybrid input-output (HIO) iterations was repeated three times followed by 20 ER iterations, for a total of 620 iterations. This algorithm was applied to each CXDP padded to a size of $256 \times 256 \times 176$ voxels. After phase retrieval, each object was mapped into orthogonal sample space and the strain was calculated using Eq. \ref{eq:strain} and the procedure found at the end of Appendix \ref{appendix:phase_retrieval}.

\begin{figure} \label{fig:seeding_error}
    \centering
    \includegraphics[scale=0.62]{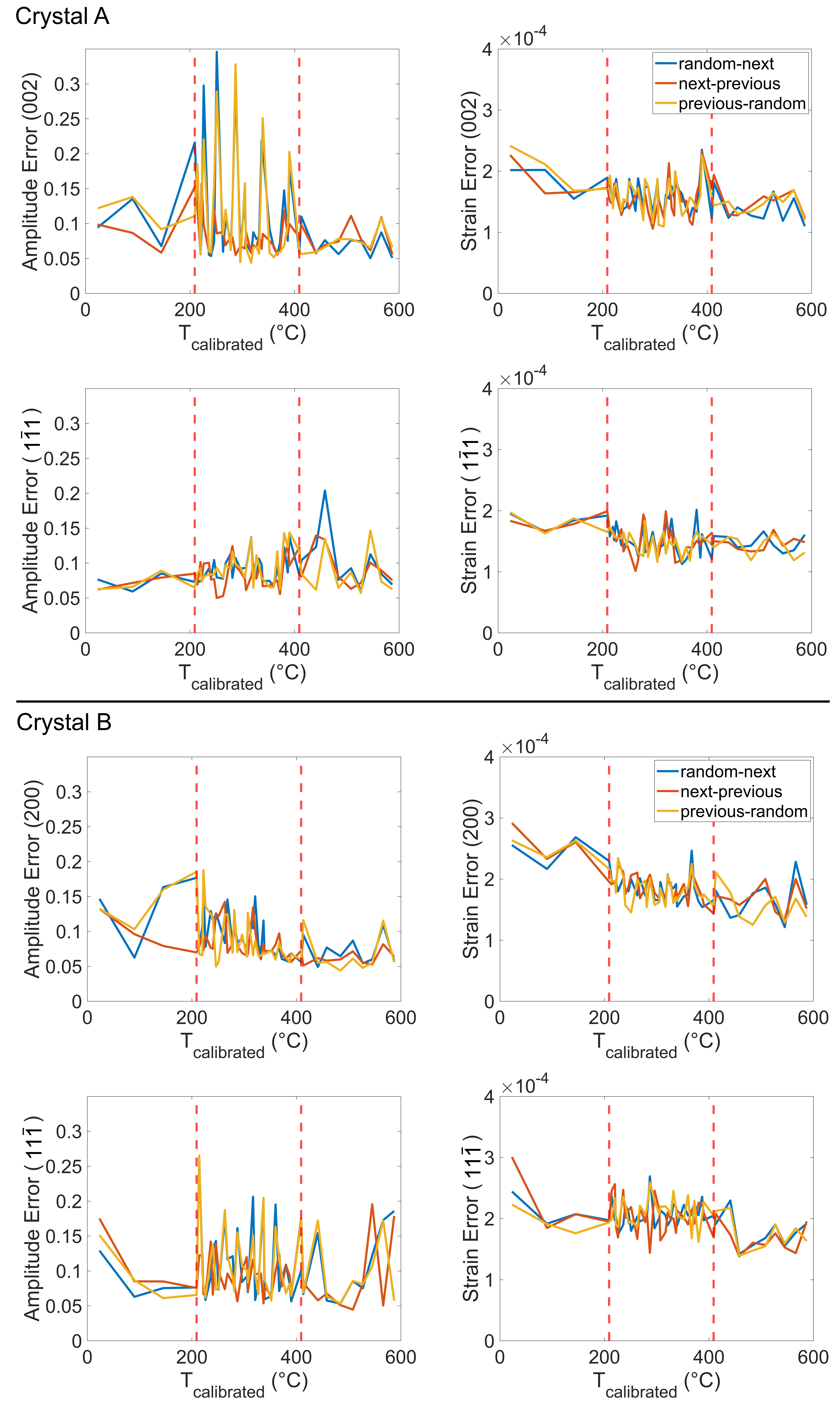}
    \caption{The amplitude and strain error, computed using Eq. \ref{eq:strain_error} and \ref{eq:amp_error} respectively, between reconstructions obtained using different seeding approaches. Each line represents the error between any two of the three seeding variables. The red error line has the lowest error on average. The large error spikes are the result of poorly converged solutions. The red dashed lines indicate where the temperature sampling changed.}
\end{figure}

To compute the error between two reconstructions generated using two different seeding approaches, we sum the difference between the two reconstructions for each voxel, $n$, and normalise it over the union of the morphologies (created using an amplitude threshold of 0.30) for that particular reflection and temperature, This yields Eq. \ref{eq:strain_error} and Eq. \ref{eq:amp_error} for the strain error and amplitude error respectively:

\begin{equation} \label{eq:strain_error}
    \mathrm{E}_{\epsilon}=\frac{\sum\limits_n|\epsilon_{\mathrm{seed\ a}}-\epsilon_{\mathrm{seed\ b}}|}{\sum\limits_n\mathrm{Mask}}
\end{equation}

\begin{equation} \label{eq:amp_error}
    \mathrm{E}_{\omega}=\frac{\sum\limits_n|\omega_{\mathrm{seed\ a}}-\omega_{\mathrm{seed\ b}}|}{\sum\limits_n\mathrm{Mask}}\\
\end{equation}

where seed a and b are any two of the three seeding scenarios listed previously. Fig. \ref{fig:seeding_error} shows how the strain and amplitude for reconstructions generated using each seeding variable compare. 

In Fig. \ref{fig:seeding_error} we observe that no two seeds generate reconstructions that are exactly the same, which would otherwise be represented by a line with negligible differences. Spikes on the plots point out that at least one reconstruction did not converge as well as the others. At temperatures where all error comparison lines overlap, the reconstructions are equally different from each other. The average strain error is $\sim$2$\times10^{-4}$, which agrees well with the uncertainty seen in previous studies \cite{Hofmann2017a,Hofmann2018a,Hofmann2020}. The red lines in Fig. \ref{fig:seeding_error} represent the error between between the two non-random seeds, and are in general lower than the error between either seed and the random guess. This shows that phasing with a seed generates reconstructions that converge more rapidly, compared to phasing with a random guess. However we cannot conclude whether phasing with a reconstruction from the next or previous temperature step yields a more accurate result. Thus, we decided for each reconstruction to be seeded with itself at the same temperature from the previous phasing round (except for the first round, where a random guess was used for all temperatures), instead of seeding with a reconstruction at a different temperature in the same round as explained in Section \ref{subsection:phasing}.

\section{Phase retrieval} \label{appendix:phase_retrieval}
The reconstruction process for every CXDP was executed in four stages, similar to the approach used by \cite{Hofmann2020}, using the output from the previous stage to seed the next phasing stage (listed below). Note the $(002)$ reflection for crystal A and $(200)$ and $(11\bar{1})$ reflections for crystal B at room temperature were padded to a size of $256 \times 256 \times 176$ voxels to make them consistent with the other CXDPs in the study. 

\begin{enumerate}
\item Each CXDP was seeded with a random guess. A guided phasing approach with 40 individuals and four generations was used \cite{Chen2007}. The first and second generations used low resolution data while the remaining generations used full resolution data. For each generation, a block of 20 ER and 180 HIO iterations, with $\beta = 0.9$, was repeated three times. This was followed by 20 ER iterations, which were averaged to return the final object. The best reconstruction was determined using a sharpness criterion, as it is the most appropriate metric for crystals containing defects \cite{Clark2015,Ulvestad2017b}. The shrinkwrap algorithm \cite{Marchesini2003} was used to periodically update the real-space support.

\item The reconstruction was seeded with the output from stage 1. This stage was identical to the previous except now we account for the longitudinal and transverse partial coherence \cite{Clark2012} using the Richardson-Lucy deconvolution algorithm.

\item The reconstruction was seeded with the output from stage 2. A block of 20 ER and 180 HIO iterations was repeated 15 times, followed by 1000 iterations of ER. The final 50 ER iterations were averaged to produce the final output. The same partial coherence settings in stage 2 were included, but guided phasing was omitted. 

\item Stage 3 was repeated, using the output from stage 3 as the seed. 



\end{enumerate}

Next, phase offsets of 0, $-\frac{\pi}{2}$, and $\frac{\pi}{2}$ were applied to the electron density for each reconstruction and the resulting phases outside $- \pi$ to $\pi$ were returned to this range by adding or subtracting $2\pi$. The resulting reconstructions for each of the three offsets were transformed from detector conjugated space to orthogonal sample space with a voxel size of $5 \times 5 \times 5 \mathrm{\ nm^3}$. Phase ramps were eliminated by centring the Fourier transform of the complex electron density. This set of phase offset data was differentiated and the minimum gradient for each voxel was used to calculate the strain, preserving its sign, to avoid discontinuities which otherwise arise at dislocation cores \cite{Hofmann2020}.  

\section{Spatial resolution} \label{appendix:spatial_resolution}
\begin{figure} \label{fig:spatial_resolution}
    \centering
    \includegraphics[width=\textwidth]{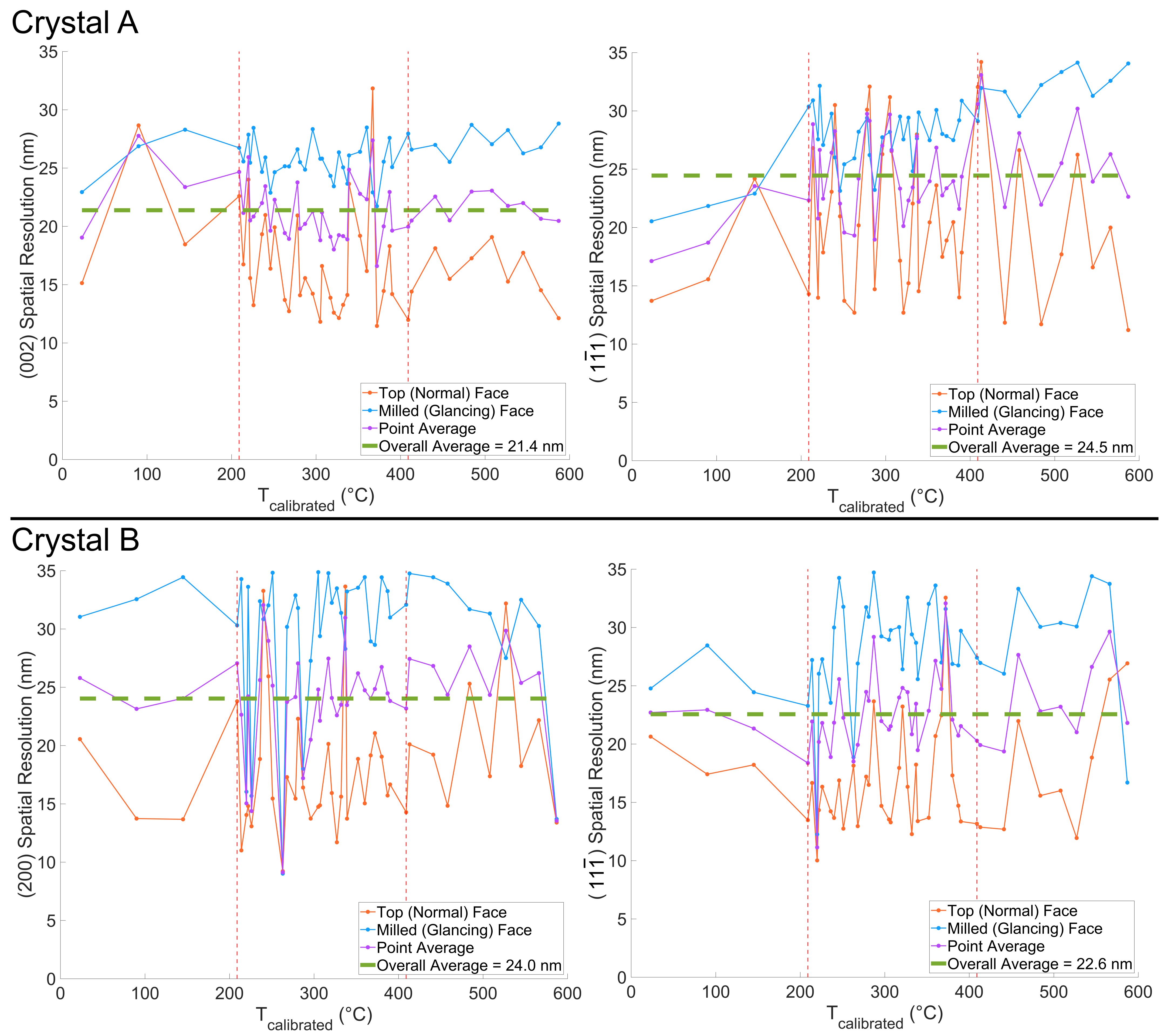} 
    \caption{Spatial resolutions with respect to temperature for each reflection. The overall average for each crystal is the mean of the point averages across all temperatures. The spatial resolution for this experiment, 23 nm, is the overall average for all reflections. The red dashed lines indicate where the temperature sampling changed.}
\end{figure}

We note that the spatial resolution is dependent on the measurement directions in reciprocal space and the geometry of the sample. The spatial resolution was calculated by considering two line profiles that cross the air-sample interface for each reflection: one normal to the top face and the other normal to the milled face. The spatial resolution was determined by differentiating the profiles and fitting the resulting peak with a Gaussian. The average $2\sigma$ of the fitted Gaussian at each temperature step is reported as the spatial resolution. A plot of the spatial resolution vs. temperature is shown in Fig. \ref{fig:spatial_resolution}.

\section{Dislocations in crystal B, (200) reflection} \label{appendix:200_phase}
\begin{figure} \label{fig:200_phase}
    \centering
    \includegraphics[scale=0.6]{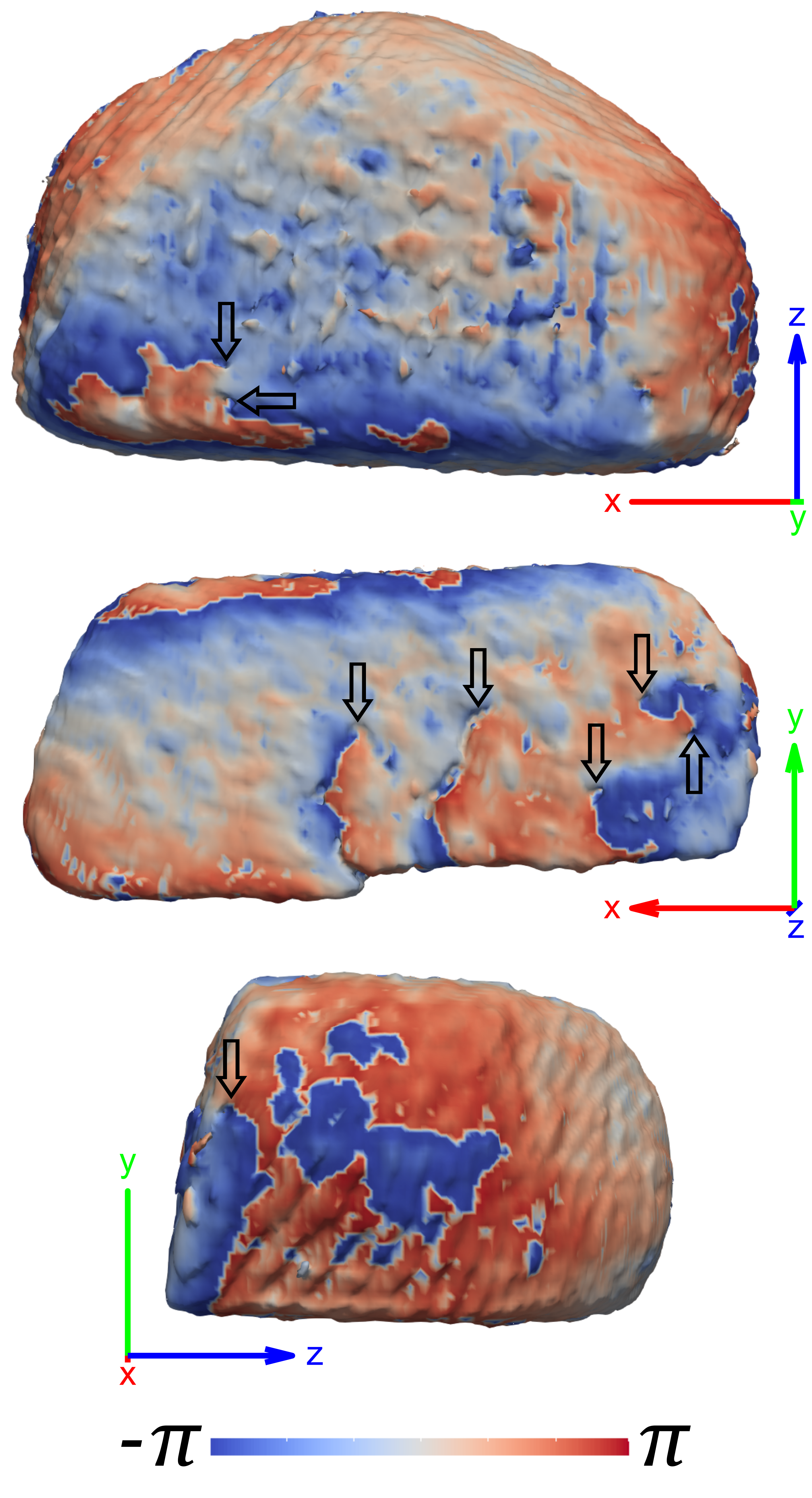} 
    \caption{A phase isosurface rendering of crystal B's $200$ reflection at room temperature, showing phase vortices reminiscent of dislocations marked by black arrows. The coordinate axes are positioned in sample space and plotted with a length of 200 nm. Here the amplitude threshold is 0.04 and no phase unwrapping was performed.}
\end{figure}

The phase for crystal B's $200$ reflection at room temperature is shown in Fig. \ref{fig:200_phase}. Vortex-like phase features that resemble dislocations are highlighted. These features are present in the trench-like regions shown in Fig. \ref{fig:strain_23_587}.

\section{Standard deviation of average strain} \label{appendix:std_avg_strain}
\begin{figure} \label{fig:std_avg_strain}
    \centering
    \includegraphics[width=\textwidth]{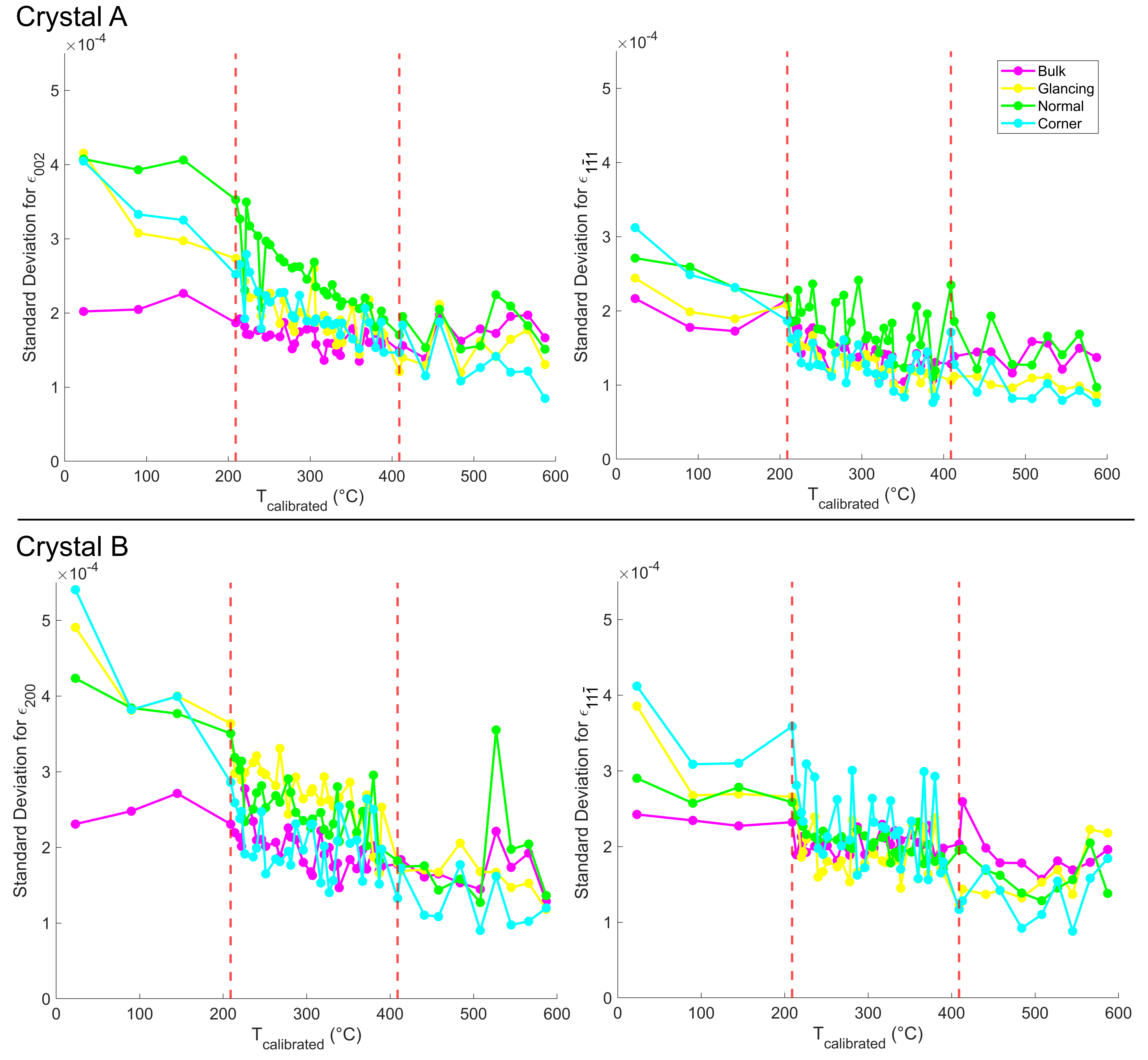} 
    \caption{Standard deviation of strain for different ROIs at each temperature point. As annealing proceeds, the standard deviation follows a decreasing trend in all ROIs, which shows a reduction in strain heterogeneity across all reflections. After $\sim$410\degree C, all standard deviation values are fairly constant, suggesting that FIB-induced strain has been removed.}
\end{figure}

The standard deviation of the strain for a particular ROI provides information about strain heterogeneity within the ROI. This can be used to evaluate the extent of annealing, whereby a lowered standard deviation corresponds to a more relaxed crystal lattice. The strain standard deviation vs. temperature for all ROIs is plotted in Fig. \ref{fig:std_avg_strain}.

\section{CXDP comparison} \label{appendix:CXDP_comparison}
\begin{figure} \label{fig:CXDP_comparison}
    \centering
    \includegraphics[scale=0.6]{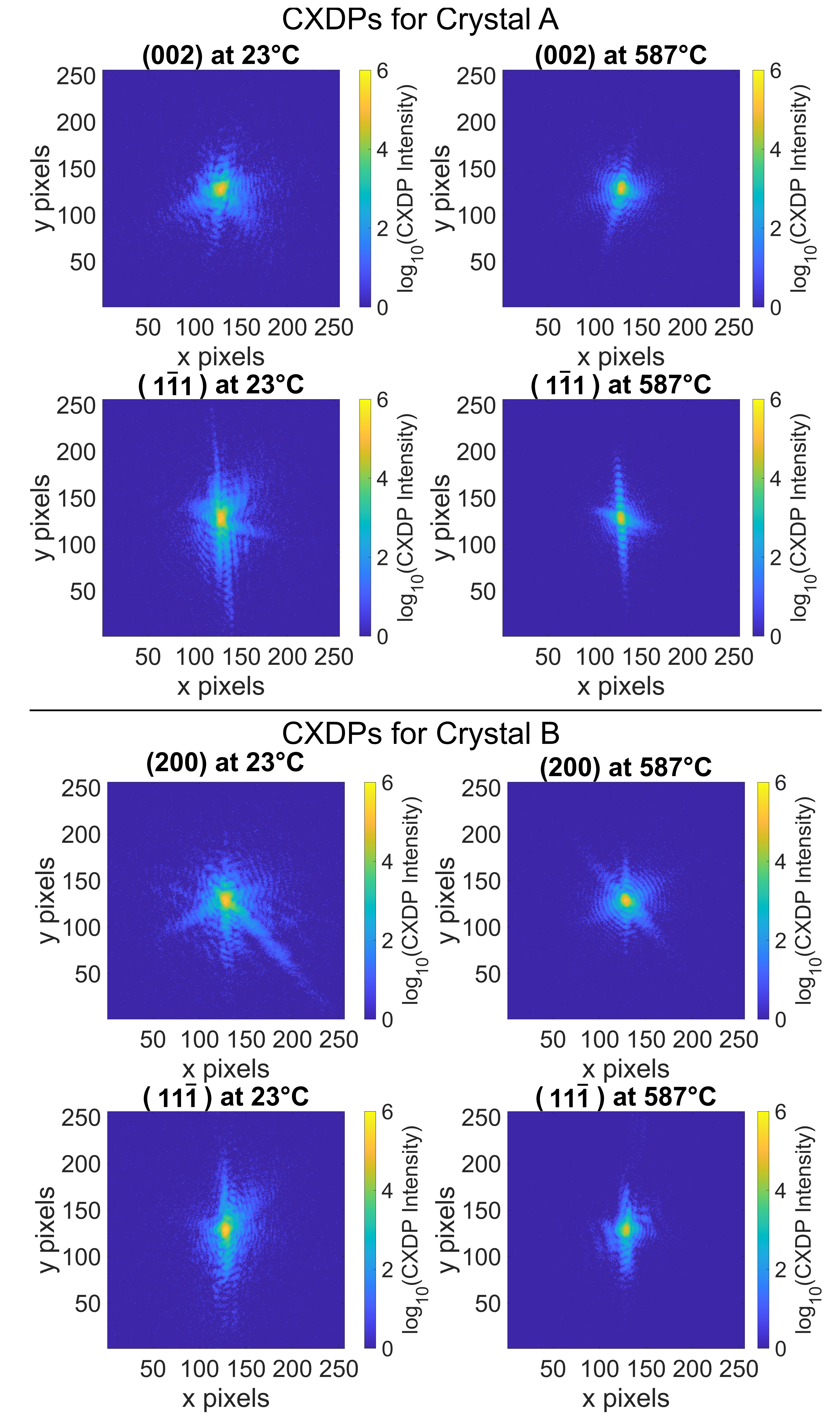} 
    \caption{Central slice of the unprocessed CXDPs at the lowest and highest temperatures for each reflection. The asymmetry in the CXDP slices before annealing indicates a high degree of strain. This strain is alleviated after heat treatment, which is demonstrated by the centrosymmetric fringes at 587\degree C. The evolution of the raw CXDPs for each reflection during heat treatment can be seen in supplementary videos SV4 and SV5.}
\end{figure}

Differences in the collected CXDPs at different temperatures show evidence of structural change. The asymmetry in the Bragg peak is indicative of strain within the crystal lattice. A comparison of the CXDPs at 23\degree C and 587\degree C is shown in Fig. \ref{fig:CXDP_comparison}.

\section{Supplementary video descriptions} \label{appendix:sv_descriptions}
Below is a list of brief descriptions for supplementary videos.
\begin{itemize}
\item SV1 shows the evolution of the surface morphology and internal strain for crystal A based on Fig. \ref{fig:strain_23_587}.
\item SV2 shows the evolution of the surface morphology and internal strain for crystal B based on Fig. \ref{fig:strain_23_587}.
\item SV3 shows facet area evolution as temperature increases for both crystals based on Fig. \ref{fig:facet_directions} and  Fig. \ref{fig:facet_sizes}.
\item SV4 shows the evolution of the CXDP slices for crystal A based on Fig. \ref{fig:CXDP_comparison}.
\item SV5 shows the evolution of the CXDP slices for crystal B based on Fig. \ref{fig:CXDP_comparison}.
\end{itemize}
\


\ack{\textbf{Acknowledgements}} \label{acknowledgements}
\sloppy The authors would like to thank Ian K. Robinson for providing the patterned Au microcrystals on silicon substrates and Ruqing Xu and Wenjun Liu at 34-ID-E for help with Laue diffraction measurements. The furnace enclosure was originally designed by Evan R. Maxey and Wonsuk Cha and the solid angle enhancement and confocal compatibility was updated by Nicholas W. Phillips. D.Y, N.W.P and F.H. acknowledge funding from the European Research Council under the European Union's Horizon 2020 research and innovation programme (grant agreement No 714697). K.S. acknowledges funding from the General Sir John Monash Foundation. The authors acknowledge use of characterisation facilities at the David Cockayne Centre for Electron Microscopy, Department of Materials, University of Oxford and use of the Advanced Research Computing (ARC) facility at the University of Oxford \cite{Richards2015}. X-ray diffraction experiments were performed at the Advanced Photon Source, a US Department of Energy (DOE) Office of Science User Facility operated for the DOE Office of Science by Argonne National Laboratory under Contract No. DE-AC02-06CH11357.
\pagebreak

\referencelist{iucr}





\end{document}